\documentclass[prb,twocolumn,superscriptaddress,longbibliography]{revtex4-2}
\usepackage{diagbox} 
\usepackage{graphicx}
\usepackage{pstricks}
\usepackage{pst-node}
\usepackage{pstricks-add}
\usepackage{amsfonts,amssymb,amsmath}
\usepackage[colorlinks=true,citecolor=green,linkcolor=red,urlcolor=blue]{hyperref}
\usepackage{subfigure}
\usepackage{calligra}
\usepackage{verbatim}

\usepackage[greek,english]{babel}

\usepackage{bbold}

\usepackage{dsfont}

\newcommand{\beq}{\begin{equation}}
\newcommand{\be}{\begin{equation}}
\newcommand{\beqn}{\begin{eqnarray}}
\newcommand{\eeq}{\end{equation}}
\newcommand{\ee}{\end{equation}}
\newcommand{\eeqn}{\end{eqnarray}}
\newcommand{\nn}{\nonumber}

\newcommand{\bem}{\begin{pmatrix}}
\newcommand{\eem}{\end{pmatrix}}

\newlength{\ldag}
\newcommand{\ket}[1]{|#1\rangle}

\usepackage{xcolor}

\begin{document}

\title{Coupled Majorana modes in  
a dual vortex of the Kitaev honeycomb model}

\author{Surajit Basak}
\email{surajit.basak@saha.ac.in}
\affiliation{Theory Division, Saha Institute of Nuclear Physics, 1/AF, Bidhannagar, Kolkata 700064, India}
\affiliation{Institute of Nuclear Physics, Polish Academy of Sciences, W. E. Radzikowskiego 152, PL-31342 Krak\'{o}w, Poland}

\author{Jean-No\"el Fuchs}
\email{jean-noel.fuchs@sorbonne-universite.fr}
\affiliation{Sorbonne Universit\'e, CNRS, Laboratoire de Physique Th\'eorique de la Mati\`ere Condens\'ee, F-75005 Paris, France}
\affiliation{Universit\'e Paris-Saclay, CNRS, Laboratoire de Physique des Solides, F-91405 Orsay, France}

\begin{abstract}
The Kitaev model is exactly solvable in terms of Majorana fermions hopping on a honeycomb lattice and coupled to a static $\mathbb{Z}_2$ gauge field, giving the possibility of $\pi$-vortices in hexagonal plaquettes. In the vortex-full sector and in the presence of a time-reversal-breaking three-spin term of strength $\kappa$, the energy spectrum is gapped and the ground state possesses an even Chern number. An isolated vortex-free plaquette acts as a ``dual vortex'' and binds a fermionic mode at finite energy $\epsilon$ in the bulk gap. This mode is equivalent to two coupled Majorana zero modes located on the same dual vortex. In a continuum approximation, we analytically compute the Majorana wavefunctions and their coupling $\epsilon$ in the two limits of small or large $\kappa$. The analytical approach is confirmed by numerical perturbation theory directly on the lattice. The latter is in excellent agreement with the full numerics on a finite-size system. We contrast our results with states bound to an isolated vortex in a topological superconductor with even Chern number.
\end{abstract}
\date{\today}

\maketitle


\section{Introduction and motivation}
The Kitaev honeycomb model~\cite{Kitaev06} emerged as a first example of an exactly solvable model for chiral quantum spin liquid, hosting fractionalized excitations and emergent gauge field. It was originally proposed as a model for spins $1/2$ on a two-dimensional (2D) honeycomb lattice, that depends on three nearest-neighbor exchange couplings $J_x$, $J_y$, $J_z$. The model can be solved exactly by mapping it onto a model for Majorana fermions hopping on the honeycomb lattice and in the presence of a static $\mathbb{Z}_2$ gauge field. The latter offers the possibility in each hexagonal plaquette to have either a zero flux (i.e. no vortex) or a $\pi$ flux (in which case, we say that there is a vortex in this plaquette). These Ising or $\mathbb{Z}_2$ vortices are also called visons~\cite{panigrahi.coleman.23, joy.rosch.22}.
The Majorana problem needs to be solved in all possible vortex sectors and to be projected onto the gauge-invariant subspace to be equivalent to the initial spin problem~\footnote{Also, the fermionic parity needs to be fixed. This is a subtle issue, see the discussions in~\cite{Pedrocchi11,Zschocke15}.}. In the isotropic limit in which $J_x=J_y=J_z=J$, and in the vortex-free sector in which the groundstate lies, the energy spectrum is gapless with a dispersion relation identical to that of graphene, i.e. with two gapless Dirac cones. Adding a small three-spin term $\kappa$ that breaks time-reversal symmetry, akin to Haldane's model~\cite{haldane.88}, a gap opens in the two Dirac cones and the system is equivalent to a topological (mean-field) superconductor with unit Chern number.

After the pioneering works of Jackeli and Khaliullin~\cite{khaliullin.05, jackeli.khaliullin.09}, several materials, such as, Na$_{2}$IrO$_{3}$~\cite{singh.manni.12, singh.gegenwart.10, revelli.sala.20}, $\alpha$-Li$_{2}$IrO$_{3}$~\cite{takayama.kato.15, williams.johnson.16}, $\alpha$-RuCl$_{3}$~\cite{plumb.clancy.14, sandilands.tian.15, zhou.li.16, nasu.knolle.16} were proposed to be described by Kitaev-like physics.
Beyond the honeycomb lattice, Kitaev model has been generalized to other trivalent graphs revealing a remarkable robustness of topological order and fractionalization. Examples range from Archimedean lattices such as the star lattice~\cite{yao.kievelson.07}, square-octagon lattice~\cite{hickey.gohlke.21}, to non-Archimedean lattices~\cite{peri.ok.20}. More recently, Kitaev physics has been explored on  hyperbolic lattices~\cite{dusel.hofman.25, mosseri.iqbal.25, vidal.mosseri.25}, on quasicrystals~\cite{keskiner.erten.23,Kim24} and on amorphous systems~\cite{Cassella23,Grushin23}, demonstrating that spin-liquid behaviour persists even in the absence of translational invariance.
Extensions to three dimensional lattices~\cite{obrien.hermanns.16, trebst.hickey.22}, such as site-depleted cubic lattice~\cite{mandal.surendran.09}, hyper-nonagon lattice~\cite{mishchenko.kato.20} further underscore the versatility of key Kitaev physics beyond two dimensions.
Parallel to these geometric generalizations, substantial effort has been devoted to understand the physics of vortices and vortex lattices in Kitaev's model~\cite{Lahtinen08,Lahtinen10,Lahtinen12,Fuchs20}. Including development of effective models for dense vortex lattices~\cite{Alspaugh24}, analyzing low energy excitations in periodically arranged vortex configurations~\cite{koga.murakami.21}, studying the charge profile around the vortices~\cite{freitas.bauer.24}.

\begin{figure}[h]
\includegraphics[width=0.9\columnwidth]{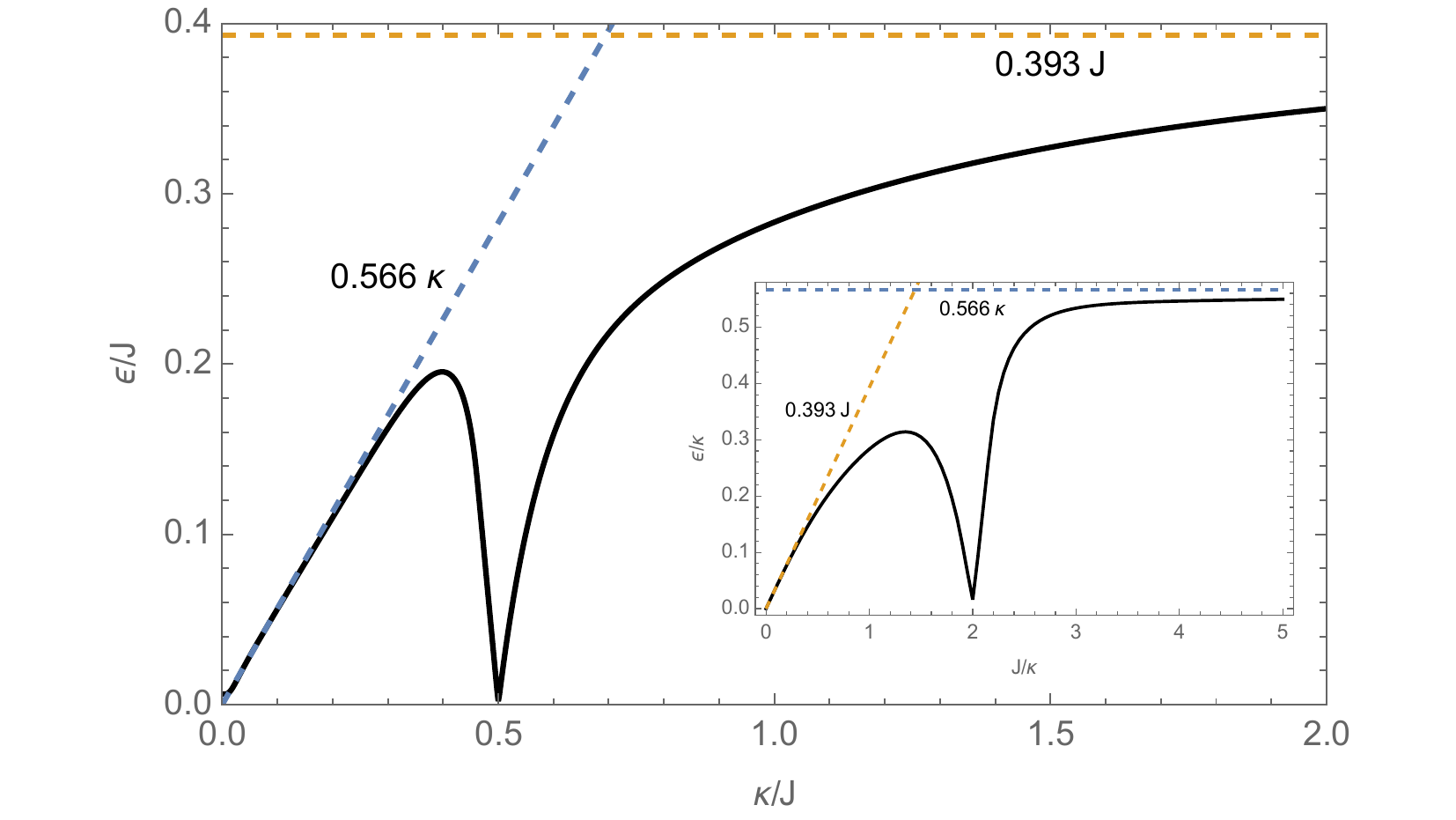}
\caption{Half-splitting $\epsilon$ (full line) as a function of $\kappa$ (both in units of $J$). The two limiting behaviors at small $\kappa$ ($\epsilon\simeq 0.566\, \kappa$) and small $J$ ($\epsilon\simeq 0.393\,  J$) are indicated as dashed lines, see also Fig. 6 in~\cite{Fuchs20}. The inset shows $\epsilon/\kappa$ versus $J/\kappa$.}
\label{fig:halfsplitting}
\end{figure}
It is well-known that an isolated vortex in a 2D topological superconductor with an odd Chern number traps an unpaired Majorana zero mode (MZM)~\cite{Volovik99,Bernevig_book}. Recently, Otten et al. have studied the case of an isolated vortex of the Kitaev model in the vortex-free sector in the presence of a small $\kappa$ term~\cite{Otten19}. In the continuum approximation, they have found an analytical expression of the wave function for the MZM that is bound to the vortex. In addition, they have also computed analytically the energy splitting between the Majorana modes in two different vortices separated by a given distance and recovered results that were known numerically~\cite{Lahtinen12} (see also~\cite{Cheng09}).

The vortex-full sector (i.e. a $\pi$-flux in every hexagonal plaquette) has been studied by Lahtinen et al.~\cite{Lahtinen08,Lahtinen10} in the small $\kappa$ limit and other values of $\kappa$ were considered in~\cite{Fuchs20}. At small $\kappa$, the dispersion relation consists of 4 gapped Dirac cones and the Chern number is $+2$. At $\kappa=J/2$, the band gap vanishes and reopens when $\kappa>J/2$ with a Chern number of $-2$. The case of a dual vortex (i.e. the absence of a vortex in an isolated plaquette within a vortex-full background) was recently investigated in the Kitaev model~\cite{Fuchs20}. It was found numerically that a dual vortex traps a fermionic mode (or a pair of Majorana modes) at finite energy $\epsilon$ for almost any $\kappa$. These authors have numerically computed the energy half-splitting $\epsilon$ as a function of $\kappa$ at fixed $J$, see Fig.~\ref{fig:halfsplitting}. In particular, this energy splitting vanishes when $\kappa=0$, $\kappa=J/2$ or when $J=0$, whereas the vortex-full band gap $\Delta$ only vanishes when  $\kappa=0$ and $\kappa=J/2$, but remains finite $\Delta=2\sqrt{3}\kappa$ when $J=0$.

Such a bound state is reminiscent of what happens in a 3D type II superconductor. There, a single superconducting vortex traps a ladder of fermionic states, as was found long ago by Caroli, de Gennes and Matricon~\cite{Caroli64}. The lowest lying states being at finite energy. It is also known that Berry phase effects are able to shift the ladder and reveal zero energy states. For example, a superconducting vortex in a 2D topological (chiral) superconductor with odd Chern number traps a MZM. In contrast, if the Chern number is even, there are mid-gap states at low but finite energy $\pm \epsilon$. This corresponds to a complex fermionic mode at finite energy or equivalently to a pair of coupled Majorana modes. 

Inspired by these two recent works~\cite{Otten19,Fuchs20}, we make a detailed study of the Majorana modes trapped in an isolated dual vortex. Our goal is to separate the Hamiltonian into two different parts: one which creates the MZMs, the other which couples them and creates the finite energy splitting. In particular, in the two limits of small $\kappa$ or small $J$, we decouple the Majorana modes to bring them to zero energy and to compute the corresponding MZM wavefunctions. Then, coupling the two modes again, we compute the energy splitting $2\epsilon$  in perturbation theory, analytically in a continuum approximation and numerically on the lattice. We compare those two approaches to a full numerical calculation on the lattice. Perturbation theory is expected to be valid as long as $\epsilon$ is small compared to the bulk gap: this means away from $\kappa=J/2$, see Fig. 6 in~\cite{Fuchs20}.

The article is organized as follows: we first review the Kitaev honeycomb model in the vortex-full sector, see Section~\ref{sec:review}. Then, in Section~\ref{sec:smallJ}, we consider the small $J$ limit. We first consider, the bulk band structure of the vortex-full sector. Then, we introduce a single isolated dual vortex in order to create MZMs. Eventually, we couple them and compute their splitting. We do this using three different methods: analytical perturbation theory in a continuum approximation, numerical perturbation theory directly on the lattice and full numerics. 
Next, we do the same study in the opposite limit of small $\kappa$, see Section~\ref{sec:smallkappa}. In the following section~\ref{sec:anachiralsupercond}, we draw the analogy to a chiral superconductor. Eventually, we discuss our results and give perspectives in a conclusion Section~\ref{sec:conclusion}. In Appendix~\ref{app:vortexfulldr}, we give details on the bulk vortex-full sector. In Appendices~\ref{app:smallJ} and~\ref{app:smallkappa}, we give details on derivations at small $J$ or at small $\kappa$.

\section{Vortex-full Kitaev model and dual vortices}
\label{sec:review}

Kitaev's honeycomb model was proposed~\cite{Kitaev06} as a two-dimensional quantum spin-1/2 model, in which spins are placed at the vertices of a honeycomb lattice, with bond-dependent interaction. It was shown in the original paper~\cite{Kitaev06} that this system can be represented by a quadratic Hamiltonian in the Majorana operators $c_{j}$ (placed at site $j$): 
\begin{equation}
    H = \frac{i}{4} \sum_{j,k} \hat{A}_{j k } c_{j} c_{k} \label{kitaevh}
\end{equation}
with $\{c_j,c_k\}=\delta_{j,k}$ and $c_j^\dagger=c_j$. Here, $\hat{A}_{jk}$ is defined as
\begin{eqnarray}
    \hat{A}_{j k } =  
    \begin{cases}
     2 J~\hat{u}_{jk}, \; \text{if $j$, $k$ are nearest neighbours,} \\
     2 \kappa~\hat{u}_{jl} \hat{u}_{lk}, \; \text{if $(k, l, j)$ is oriented clockwise,} \\
     0, \;  \text{otherwise.}
    \end{cases}
    \label{A}
\end{eqnarray}
$J$ is the nearest neighbor interaction strength and the part proportional to $\kappa$ comes from a three-spin term which breaks the time-reversal symmetry (TRS). Without loss of generality, we assume that $\kappa\geq 0$. $\hat{u}_{jk} = -\hat{u}_{kj}$, is an operator, defined on the link connecting sites $j$ and $k$, with eigenvalues $\pm 1$. $\hat{u}_{jk}$ commutes with the Hamiltonian, hence the Hilbert space splits up into several sectors, each labeled by the eigenvalues of $\hat{u}$. But it turns out that, $\hat{u}_{jk}$ is a $\mathbb{Z}_{2}$ gauge variable, so it is more sensible to label each non-equivalent sector of the Hamiltonian with the eigenvalues of the gauge-invariant operator 
\begin{eqnarray}
    \hat{W}_{p} = \prod_{(j,k)\in \text{boundary of }p} \hat{u}_{jk}.
\end{eqnarray}
$\hat{W}_{p}$ is an operator defined at each plaquette $p$ that commutes with the Hamiltonian and has eigenvalues $\pm 1$. In terms of the eigenvalues, one can write
\begin{eqnarray}
    w_{p} = \prod_{(j,k)\in \text{boundary of }p} u_{jk}.
\end{eqnarray}

\begin{figure}[!h]
\includegraphics[width=0.7\linewidth]{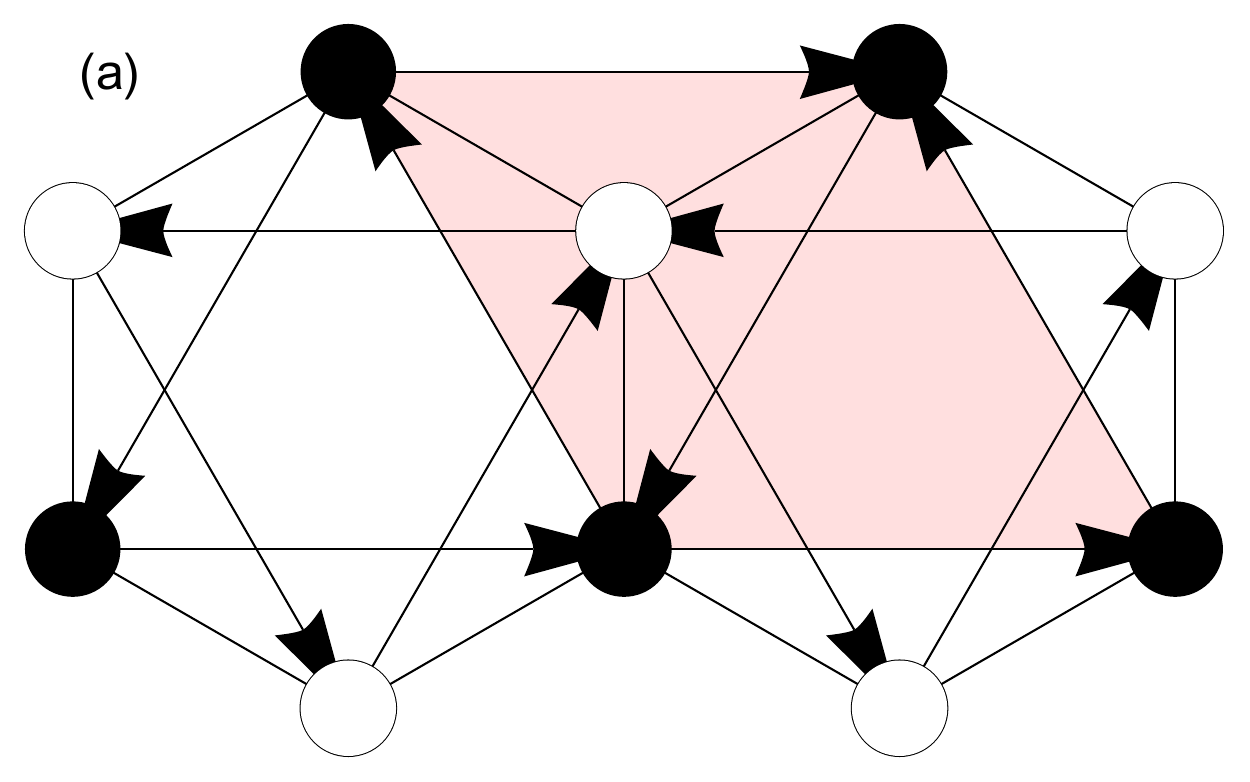}
\includegraphics[width=0.7\linewidth]{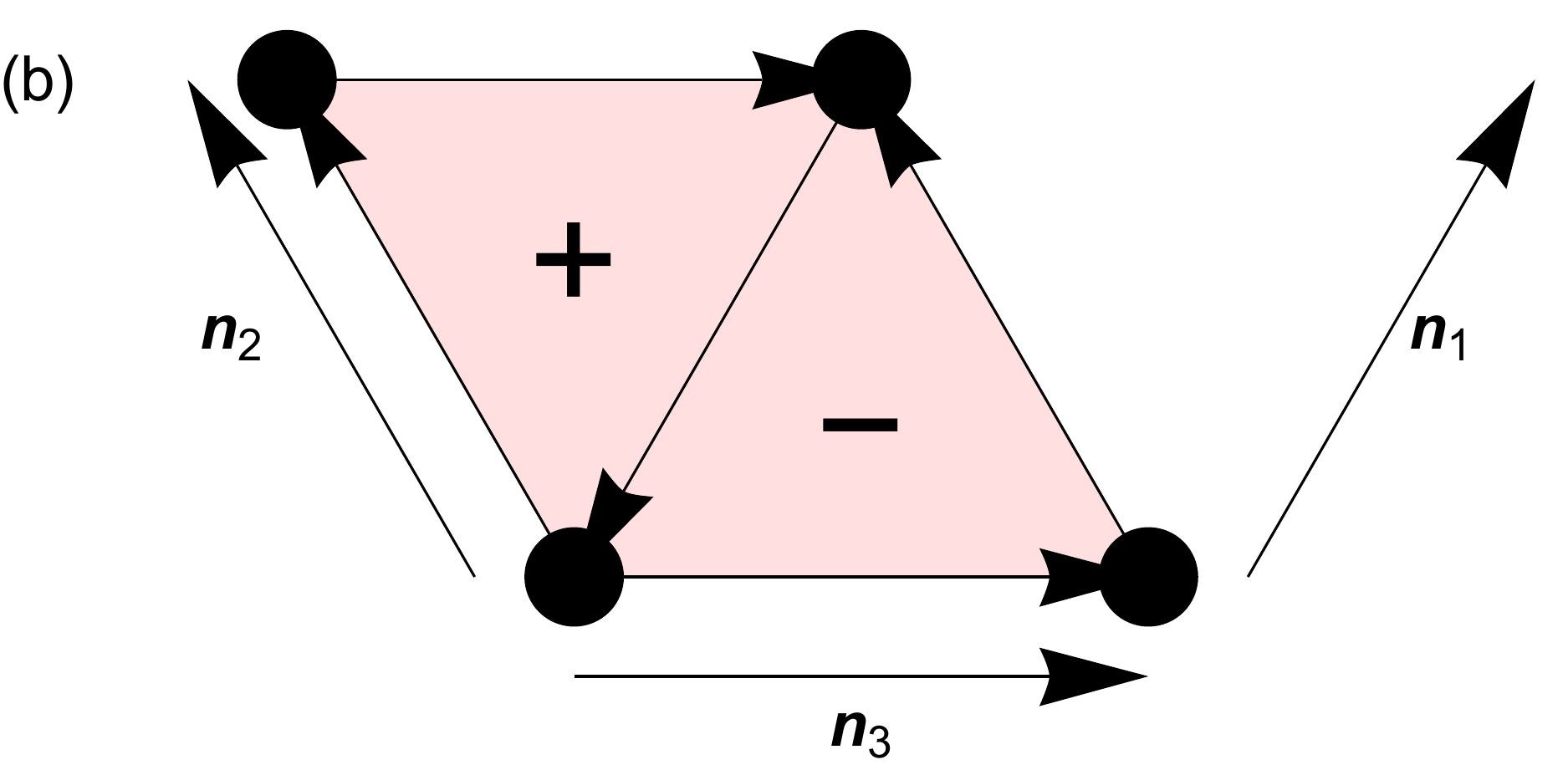}
\caption{\emph{Vortex-free sector}. (a) The Kitaev gauge (i.e. $+iJ$ hopping from black to white sites) is assumed everywhere on the nearest-neighbor links of the honeycomb lattice. 
(b) When $J=0$, the two triangular sublattices are decoupled. We show the unit cell for the triangular sublattice of black sites. The unit cell vectors are $\mathbf{n}_3$ and $\mathbf{n}_2$ ($\mathbf{n}_1$ is the remaining nearest-neighbor vectors on the triangular lattice). The signs indicate whether the flux is $\pm \pi/2$ in a triangle.
}
\label{fig.vortexfree}
\end{figure}

In this way, the original Hilbert space $\mathcal{L}$ splits up into several independent sectors, each labeled by the set $\{w_{1}, w_{2}, \cdots w_{N_p} \}$ ($N_p$ is the number of plaquettes in the system): $\mathcal{L} = \oplus_{\{w\}} \mathcal{L}_{\{w\}}$. The variable $w_{p}$ can be regarded as the magnetic flux through the plaquette $p$. A plaquette is defined to carry a vortex if $w_{p} = -1$, and no vortex if $w_p=1$. For a given sector, the Kitaev's Hamiltonian can be written as
\begin{equation}
    H = \frac{i}{4} \sum_{j,k} A_{j k } c_{j} c_{k} = H_J + H_\kappa. \label{kitaevhev}
\end{equation}
$A_{j k }$ now being a skew-symmetric matrix [cf. Eq.~\eqref{A}]. It was shown by Lieb~\cite{Lieb94} that, for $\kappa = 0$, the ground state of this model lies in the sector with $w_{p} = 1$, for all $p$. This is known as the vortex-free sector. A convenient gauge for this sector is the Kitaev gauge: 
\begin{eqnarray}
u_{jk}^\text{K} &=& + 1 \text{ if $k$ is a black site and $j$ a white site}\nn \\
&=& - 1 \text{ if $k$ is a white site and $j$ a black site},
\end{eqnarray}
where black and white denote the two triangular sublattices of the honeycomb lattice [see Fig.~\ref{fig.vortexfree}(a)]. By convention, we do not put arrows on nearest-neighbor links of the honeycomb lattice, but have in mind that they are oriented from black to white, unless the link has been flipped. In the Kitaev gauge, the next-nearest neighbor hopping are oriented such that they rotate anti-clockwise around the center of a hexagon.

In this work, we will be mainly interested in the vortex-full sector, i.e. $w_{p} = -1$ for all plaquettes. Starting from the Kitaev gauge, by flipping (i.e. $u_{jk}\to - u_{jk}$) one vertical link out of two, see the blue links in Fig.~\ref{fig:vortexfull}(a), we can realize it. Indeed, each flipped link creates a pair of vortices (i.e. $w_p=-1$) in the neighboring hexagonal plaquettes. This defines our vortex-full gauge $u_{jk}^\text{full}$. 

In a vortex-full background, by flipping one link we can introduce a pair of vortex-free plaquettes. By flipping further links along a string, we can separated the two vortex-free plaquettes. An isolated vortex-free plaquette in a vortex-full background is thought of as being a ``dual vortex". We call $u_{jk}^\text{dual}$ the gauge that realizes two isolated dual vortices that are far apart. In the numerical calculations, we will work on a closed surface of finite size and always have two dual vortices. However, in the analytical calculation, we will work in an infinite system in which we concentrate near one end of the string and send the other end to infinity. Such a half-infinite string of flipped links is therefore attached to a single isolated dual vortex.

According to the tenfold periodic table classification of free fermions~\cite{schnyder.ryu.08, Kitaev09}, Kitaev's Hamiltonian (\ref{kitaevhev}) belongs to the D-class of topological superconductors. TRS is broken as soon as $\kappa\neq 0$. In addition, the model can be mapped into a free Majorana fermion problem (\ref{kitaevhev}), which shows that the particle-hole symmetry (PHS) is present and such that it squares to +1. 
Members of the D-class are characterized by a Chern number $\nu$. 
Although the classification of free-fermion topological superconductors is $\mathbb{Z}$, the classification of the corresponding interacting bulk anyon theories is reduced to $\mathbb{Z}_{16}$, which is known as Kitaev's sixteen-fold way~\cite{Kitaev06}.
The vortex-full sector has $\nu=+2$, when $0<\kappa<J/2$, and $\nu=-2$ when $\kappa>J/2$~\cite{Lahtinen10,Fuchs20}. When $\kappa=0$ or $\kappa=J/2$, the gap vanishes and the Chern number is not defined. By the analogy to a topological superconductor with even Chern number, such a dual vortex is expected to trap bound states at finite energy (and not a MZM)~\cite{Volovik99}.

In the following, we study the mid-gap states bound to a dual vortex in the $\kappa\gg J$ and $\kappa \ll J$ limits. We start with the simplest case of small $J$. 

\section{Small $J$ limit}
\label{sec:smallJ} 
In this section, we consider the limit $J\ll \kappa$. Starting from the vortex-full gauge $u^\text{full}$, flipping a half-infinite string of links creates an isolated dual vortex [see Fig.~\ref{fig.carlo_vortex}(a)] corresponding to the gauge $u^\text{dual}$. The Hamiltonian
\beqn
H = H_{\kappa}[u^\text{dual}]+ H_{J}[u^\text{dual}]
\label{eq:totham2}
\eeqn
can be split in four terms:
\beqn
H =  H_{\kappa\text{-bulk}} + H_{\kappa\text{-string}} + H_{J\text{-bulk}} + H_{J\text{-string}},
\label{eq:totham}
\eeqn
by defining $H_{\kappa\text{-bulk}} = H_{\kappa}[u^\text{full}]$, $H_{\kappa\text{-string}} = H_{\kappa}[u^\text{dual}]-H_{\kappa}[u^\text{full}]$, $H_{J\text{-bulk}} = H_{J}[u^\text{full}]$, $H_{J\text{-string}} = H_{J}[u^\text{dual}]-H_{J}[u^\text{full}]$. 
In section~\ref{sec:anasmallJ}, we do analytical perturbation theory in a continuum approximation and consider the different terms in succession: we first study $H_{\kappa\text{-bulk}}$ that produces a gapped band structure (section \ref{sec:decoupled}), then add $H_{\kappa\text{-string}}$ to create a single dual vortex that traps two decoupled MZMs (section \ref{subsec:smallJ_Hlink}) and eventually add the $H_{J}=H_{J\text{-bulk}} + H_{J\text{-string}}$ that couples the Majorana modes (section \ref{sec:energysplitting}). We then check our analytical results by performing numerical perturbation theory directly on the lattice (section~\ref{sec:nptsmallJ}) and a fully numerical calculation (section~\ref{sec:fnsmallJ}), both on a finite-size system.

\subsection{Analytic perturbation theory in the continuum}
\label{sec:anasmallJ}
\subsubsection{Decoupled triangular lattices and gapped Dirac cones} \label{sec:decoupled}
\begin{figure}[!h]
\includegraphics[width=\linewidth]{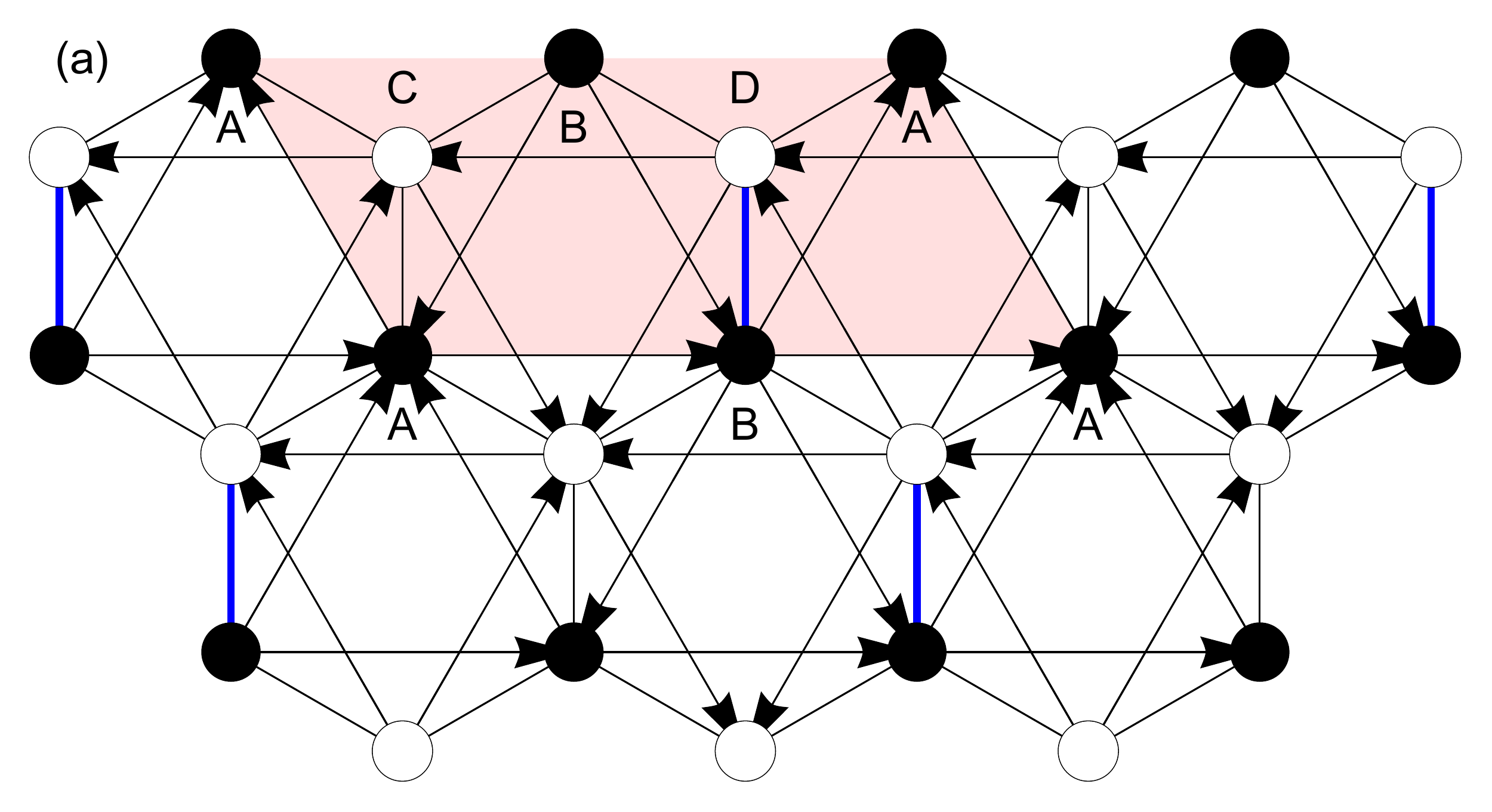}
\includegraphics[width=\linewidth]{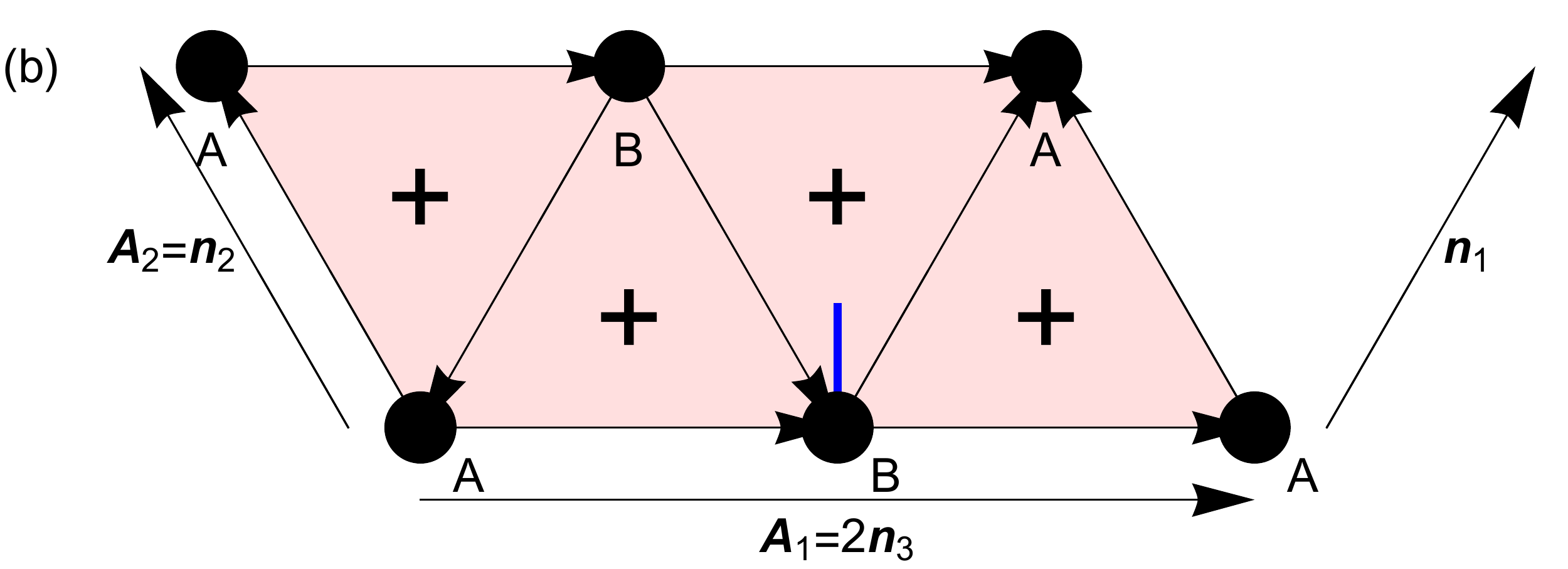}
\caption{\emph{Vortex-full sector}. (a) The Kitaev gauge (i.e. $+iJ$ hopping from black to white sites) is assumed everywhere on the nearest-neighbor links of the honeycomb lattice, except on flipped links that are shown in blue (i.e. $-iJ$ hopping from black to white sites). The unit cell is shown as a pink parallelogram and contains four sites labeled $A$, $B$, $C$ and $D$. (b) At $J=0$, the two triangular sublattices ($AB$ and $CD$) are decoupled. We show the unit cell for the triangular sublattice of black sites ($AB$). The flipped links are indicated by blue tails. The unit cell vectors are $\mathbf{A}_1$ and $\mathbf{A}_2$, and $\mathbf{n}_1$, $\mathbf{n}_2$, $\mathbf{n}_3$ are the nearest-neighbor vectors on the triangular lattice. The ``+'' signs indicate that the flux in every triangle is $+\pi/2$.}
\label{fig:vortexfull}
\end{figure}

In the $J=0$ limit, the vortex-full problem (with gauge $u_{jk}^\text{full}$) separates in two decoupled triangular lattices with $\pi/2$ flux per triangle and $\kappa$ hopping amplitude~\cite{Alspaugh24}. The black sites ($AB$) form a triangular lattice that is decoupled from the triangular lattice of white sites ($CD$), see Fig.~\ref{fig:vortexfull}. An important point to make is that, even when $J=0$, the gauge $u_{jk}$ on the $J$-links still matters as it determines the sign of the imaginary hopping on the $\kappa$-links.

The Hamiltonian in the vortex-full sector, in canonical basis~(\ref{kdual}), with $J = 0$ can be written as
\begin{eqnarray}
    H = H_{\kappa\text{-bulk}} = \sum_{\mathbf{k}} \mathbf{\Psi}^{\dagger}_{\mathbf{k}} \left(\  \begin{array}{cc} \mathcal{H}^{AB}(\mathbf{k}) & 0\\
    0 & \mathcal{H}^{CD} (\mathbf{k}) \end{array}\right) \mathbf{\Psi}_{\mathbf{k}},
\end{eqnarray}
where $\mathbf{\Psi}^{T}_{\mathbf{k}} = (\chi_{\mathbf{k}}^A,\chi_{\mathbf{k}}^B,\chi_{\mathbf{k}}^C,\chi_{\mathbf{k}}^D)^{T}$ and the reciprocal space Hamiltonians are
\begin{eqnarray}
\mathcal{H}^{AB}(\mathbf{k}) 
&=& 2 \kappa [\sin(\mathbf{k}\cdot \mathbf{n}_3) \sigma_x -\cos(\mathbf{k}\cdot \mathbf{n}_1) \sigma_y +\sin(\mathbf{k}\cdot \mathbf{n}_2) \sigma_z] \nonumber \\
    & = &\mathbf{h}^{AB}(\mathbf{k})\cdot \boldsymbol{\sigma}, \label{HAB}
\end{eqnarray}
\begin{eqnarray}
\mathcal{H}^{CD}(\mathbf{k}) 
& =& -2 \kappa [\sin(\mathbf{k}\cdot \mathbf{n}_3) \tau_x +\cos(\mathbf{k}\cdot \mathbf{n}_1) \tau_y +\sin(\mathbf{k}\cdot \mathbf{n}_2) \tau_z] \nonumber \\
    & =& \mathbf{h}^{CD}(\mathbf{k})\cdot \boldsymbol{\tau}.
\end{eqnarray}
$\sigma_{i}$ ($\tau_{i}$), with $i=x,y,z$, are the Pauli matrices in the black (white) sublattice space, and $\mathbf{n}_1=(1/2,\sqrt{3}/2)$, $\mathbf{n}_2=(-1/2,\sqrt{3}/2)$ and $\mathbf{n}_3=(1,0)$ are the nearest-neighbor vectors on the triangular lattice, see Fig.~\ref{fig.vortexfree}. 

\begin{figure}[!h]
\includegraphics[width=\linewidth]{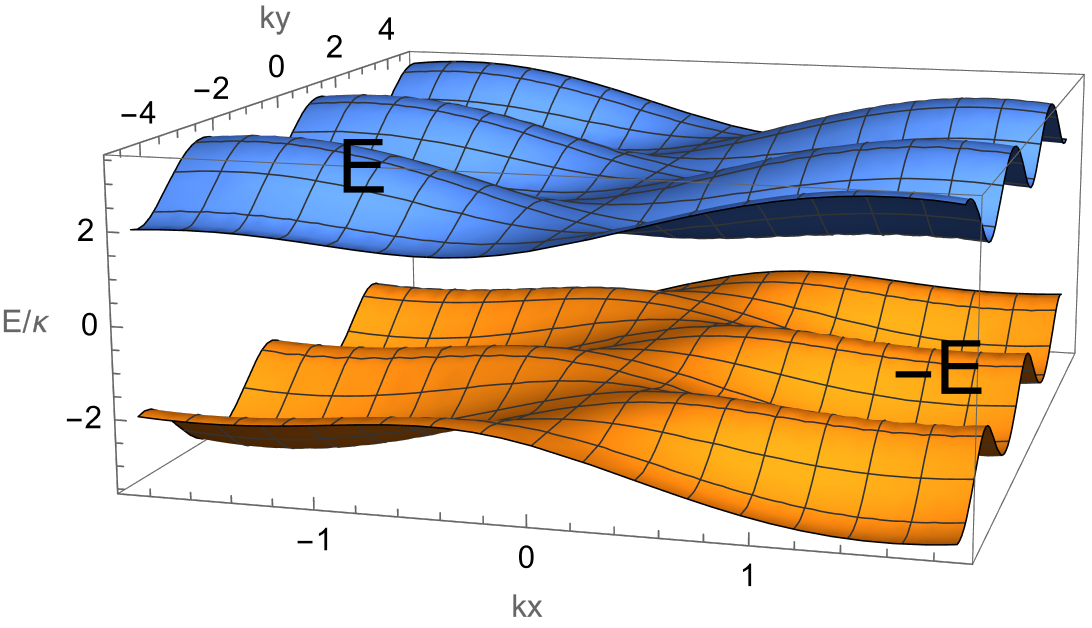}
\includegraphics[width=\linewidth]{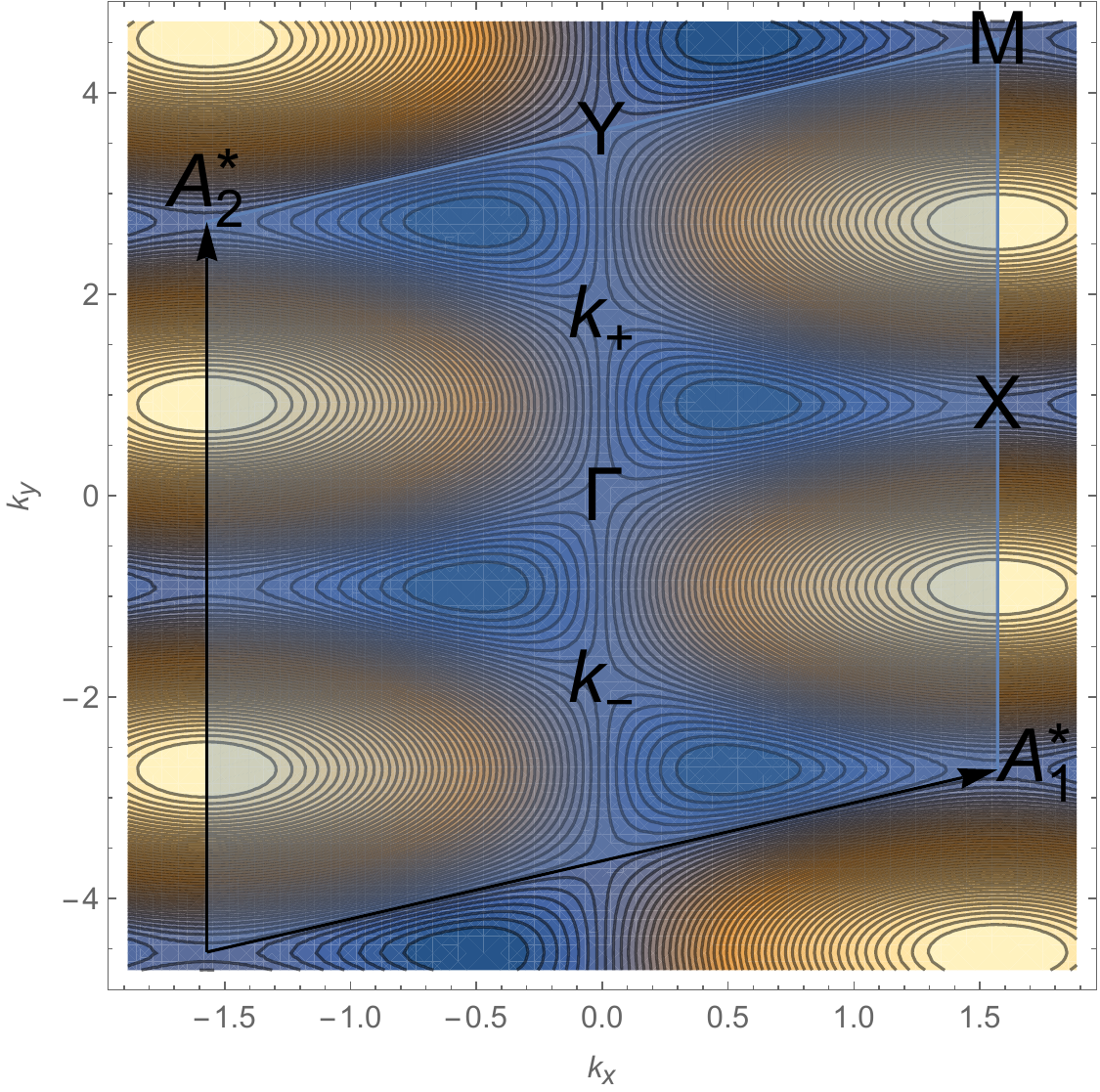}
\caption{(Top) Energy bands $\pm E(\mathbf{k})$ (in units of $\kappa$), see Eq.~\eqref{eq:disptriang}. (Bottom) Isoenergy lines of $E(\mathbf{k})$. Also indicated are the Brillouin zone, the reciprocal lattice vectors $\mathbf{A}_1^*$ and $\mathbf{A}_2^*$, the high symmetry points $\Gamma$, $X$, $Y$ and $M$ and the position of the two Dirac points $\mathbf{k}_\pm = \pm \mathbf{A}_2^*/4$. }
\label{fig.BZisoenergy}
\end{figure}

Now, we focus on the Hamiltonian $\mathcal{H}^{AB}(\mathbf{k})$ [see Fig.~\ref{fig:vortexfull}(b)], the other Hamiltonian $\mathcal{H}^{CD}(\mathbf{k})$ being very similar. This Hamiltonian represents a triangular lattice in a uniform magnetic field (at quarter-flux quantum or $+\pi/2$ per triangle), which was studied long ago by Claro and Wannier~\cite{Claro79} in the context of the Hofstadter butterfly. The Bravais lattice vectors are $\mathbf{A}_1=2 \mathbf{n}_3=(2,0)$ and $\mathbf{A}_2=\mathbf{n}_2$ and the reciprocal ones are $\mathbf{A}_1^*=(\pi,\pi/\sqrt{3})$ and $\mathbf{A}_2^*=(0,4\pi/\sqrt{3})$. The energy spectrum is made of two bands $\pm E(\mathbf{k})$ with
\begin{eqnarray}\label{eq:disptriang}
E(\mathbf{k})&=& \sqrt{f(\mathbf{k})}\\
&=&2\kappa \sqrt{\sin^2(\mathbf{k}\cdot \mathbf{n}_3)+\cos^2(\mathbf{k}\cdot \mathbf{n}_1)+\sin^2(\mathbf{k}\cdot \mathbf{n}_2)},\nn
\end{eqnarray}
see Eq.~\eqref{eq:dispvf} with $J=0$, and Fig.~\ref{fig.BZisoenergy}. 
They have a width of $\sqrt{3}\kappa$ each, are symmetric under the particle-hole transformation $E\to -E$, and are separated by a large gap of $2\sqrt{3}\kappa$. In addition to $E(\mathbf{k}+l \mathbf{A}_1^* +m \mathbf{A}_2^*)=E(\mathbf{k})$ with $l$ and $m$ integers, the dispersion relation has the remarkable symmetry $E(\mathbf{k}+\mathbf{A}_2^*/2)=E(\mathbf{k})$ (see Fig.~\ref{fig.BZisoenergy}). The lower band carries a Chern number of $\nu=-1$. The two sublattices together produce a Chern number of $\nu=-2$ (for the negative energy bands) as expected for the vortex-full case with $\kappa>J/2$. 

From the analysis of~\cite{Otten19} in the vortex-free sector, we expect that a vortex will trap a MZM, that, in reciprocal space, is localized at a pair of Dirac points. However, due to the large gap between the valence and conduction bands of the energy dispersion, it is not obvious to find the Dirac points simply by looking at the energy spectrum. However, because the bands carry non-zero Chern numbers, we know that there must be Dirac points and that they occur in pair (a fact known as ``fermion doubling'') such as to give an integer Chern number~\cite{haldane.88,Sticlet12}. How can we find these hidden Dirac fermions? We follow~\cite{Sticlet12} and arbitrarily choose to set $h^{AB}_z\to 0$ and then search for the $\mathbf{k}$-points such that $h^{AB}_x(\mathbf{k})=0$ and $h^{AB}_y(\mathbf{k})=0$~\footnote{Other choices lead to Dirac points being at $\Gamma$ and $Y$ (if $h^{AB}_y\to 0$) or at $X$ and $M$ (if $h^{AB}_x\to 0$). In all cases, these are saddle points of the dispersion relation, see Fig.~\ref{fig.BZisoenergy}.}. We find two solutions: $\mathbf{k}_{\pm}=(0,\pm\pi/\sqrt{3}) = \pm \mathbf{A}_{2}^{*}/4$, see Fig.~\ref{fig.BZisoenergy}. These are the two Dirac points. 
Because $h^{AB}_z$ is actually not zero, they are gapped with mass $m_\xi=h^{AB}_z(\mathbf{k}_\xi)$ and are also characterized by a chirality $\mathcal{C}_\xi=\text{sign }[\partial_{k_x}\mathbf{h}^{AB} (\mathbf{k}_\xi)\times \partial_{k_y} \mathbf{h}^{AB}(\mathbf{k}_\xi)]_z$, where the valley index $\xi=\pm$ refers to $\mathbf{k}_{\pm}=\pm \mathbf{k}_+$, i.e. $\mathbf{k}_\xi = \xi \mathbf{k}_+$. 
Starting from Eq.~\ref{HAB}, we therefore expand the Hamiltonian around the two valleys by writing $\textbf{k} = \textbf{k}_{\xi}+\textbf{q}$, where $|\textbf{q}|<<1$. For the '+' valley,
\begin{align*}
(\mathbf{k}_{+}+\mathbf{q})\cdot\mathbf{n}_{1} &= \frac{\pi}{2} + \mathbf{q}\cdot\mathbf{n}_{1}, \\
(\mathbf{k}_{+}+\mathbf{q})\cdot\mathbf{n}_{2} &= \frac{\pi}{2} + \mathbf{q}\cdot\mathbf{n}_{2}, \\
(\mathbf{k}_{+}+\mathbf{q})\cdot\mathbf{n}_{3} &= \mathbf{q}\cdot\mathbf{n}_{3},
\end{align*}
while for the '-' valley, one simply replaces $\frac{\pi}{2} \to -\frac{\pi}{2}$. Finally, using
$\sin x \simeq x$,
$\cos(\frac{\pi}{2}+x) \simeq -x$ and $\sin(\frac{\pi}{2}+x) \simeq 1$, to linear order in $\textbf{q}$, Eq.~\ref{HAB} becomes:
\begin{eqnarray}
\mathcal{H}^{AB}_{\xi}(\mathbf{q}) &=&\mathcal{H}^{AB}(\mathbf{k}_\xi + \mathbf{q}) \nn \\
&=& 2\kappa\left[\mathbf{q}\cdot\mathbf{n_{3}}\sigma_x + \xi \mathbf{q}\cdot\mathbf{n_{1}}\sigma_y + \xi \sigma_z\right].
\label{brillouinexp}
\end{eqnarray}
It has the form of a 2D massive Dirac Hamiltonian. The two valleys have opposite chirality $\mathcal{C}_\xi=\xi$ (i.e. relative sign between the $\mathbf{q}\cdot\mathbf{n_{1}}$ and $\mathbf{q}\cdot\mathbf{n_{3}}$ terms) and opposite mass sign, $\text{sign}(m_\xi)=\xi$, (i.e. sign of the constant term in $\sigma_z$) so that they contribute to a non-zero Chern number of the lowest band~\cite{Sticlet12}:
\begin{equation}
\nu = -\frac{1}{2} \sum_{\xi=\pm}  \mathcal{C}_\xi \text{ sign}(m_\xi) = - 1.
\end{equation}

The points $\mathbf{k}_\pm$ actually correspond to saddle points of the full dispersion relation rather than to band extrema as expected for gapped Dirac cones. The linearized Hamiltonian in Eq.~\eqref{brillouinexp} is therefore meant to capture the correct spinor wave function structure (giving rise in particular to the Chern number) at the cost of being a poor description of the energy dispersion. The reason is that, because of the even/odd effect, we know that the correct Chern numbers are mandatory in order to capture the MZM bound-state. The limitations of this approach are reflected in the quantitative deviations from the full lattice results discussed below.

In the end, the effective Hamiltonian is similar to that of a $p_x+ip_y$ superconductor with two valleys~\cite{Otten19}. 
Together with the other ($CD$) triangular lattice, we therefore have two decoupled $p_x+i p_y$ superconductors, each with two valleys. This corresponds to 4 gapped Dirac cones in total. It is well-known that a vortex in a $p_x+i p_y$ superconductor traps a single unpaired MZM~\cite{Read00,Bernevig_book}. In the following section, we add a (dual) vortex to the picture in order to create a MZM for each decoupled triangular lattice.

\subsubsection{Dual vortex and MZM wavefunctions}
\label{subsec:smallJ_Hlink} 
\begin{figure}[!h]
\includegraphics[width=\linewidth]{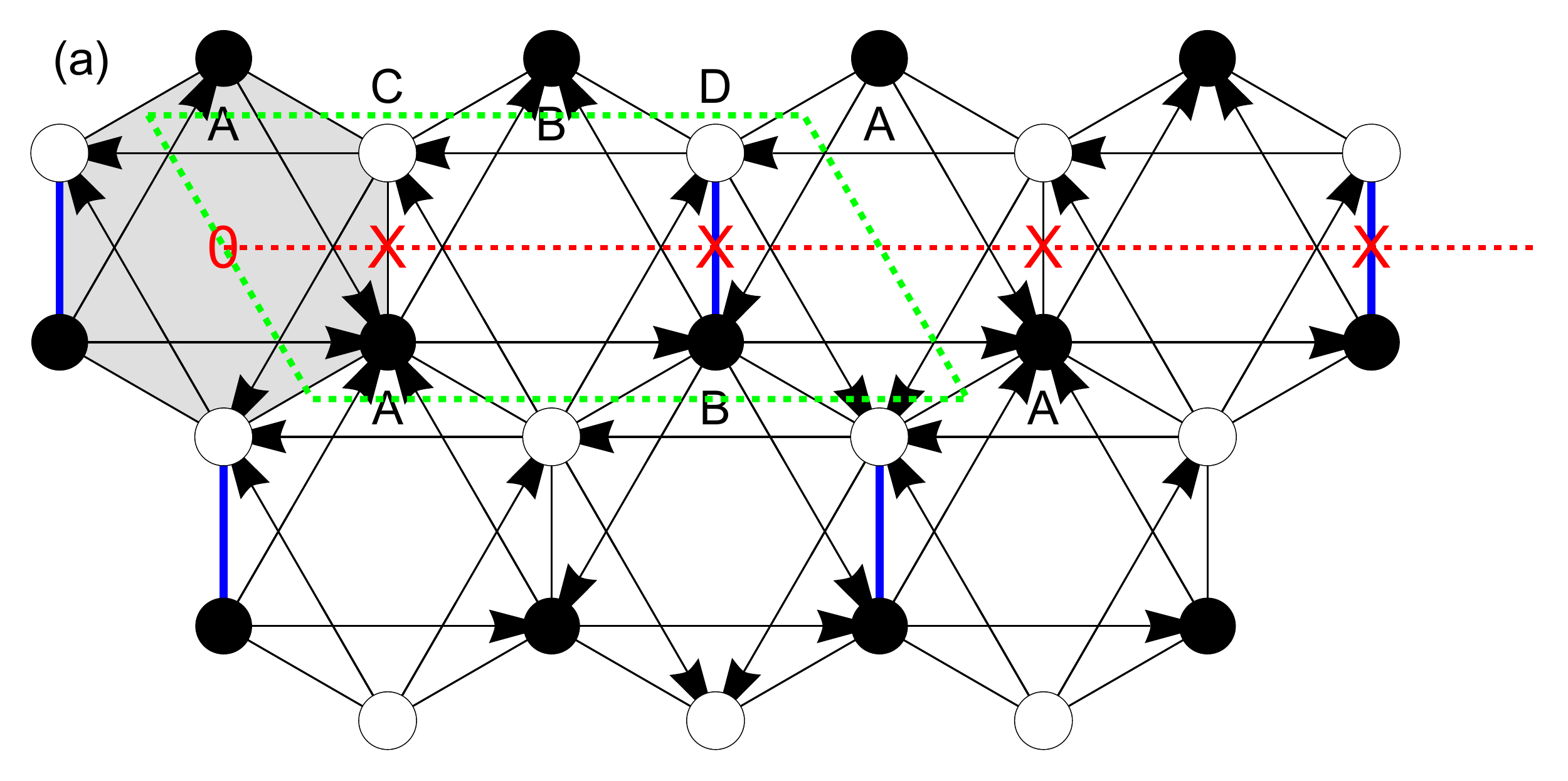}
\includegraphics[width=\linewidth]{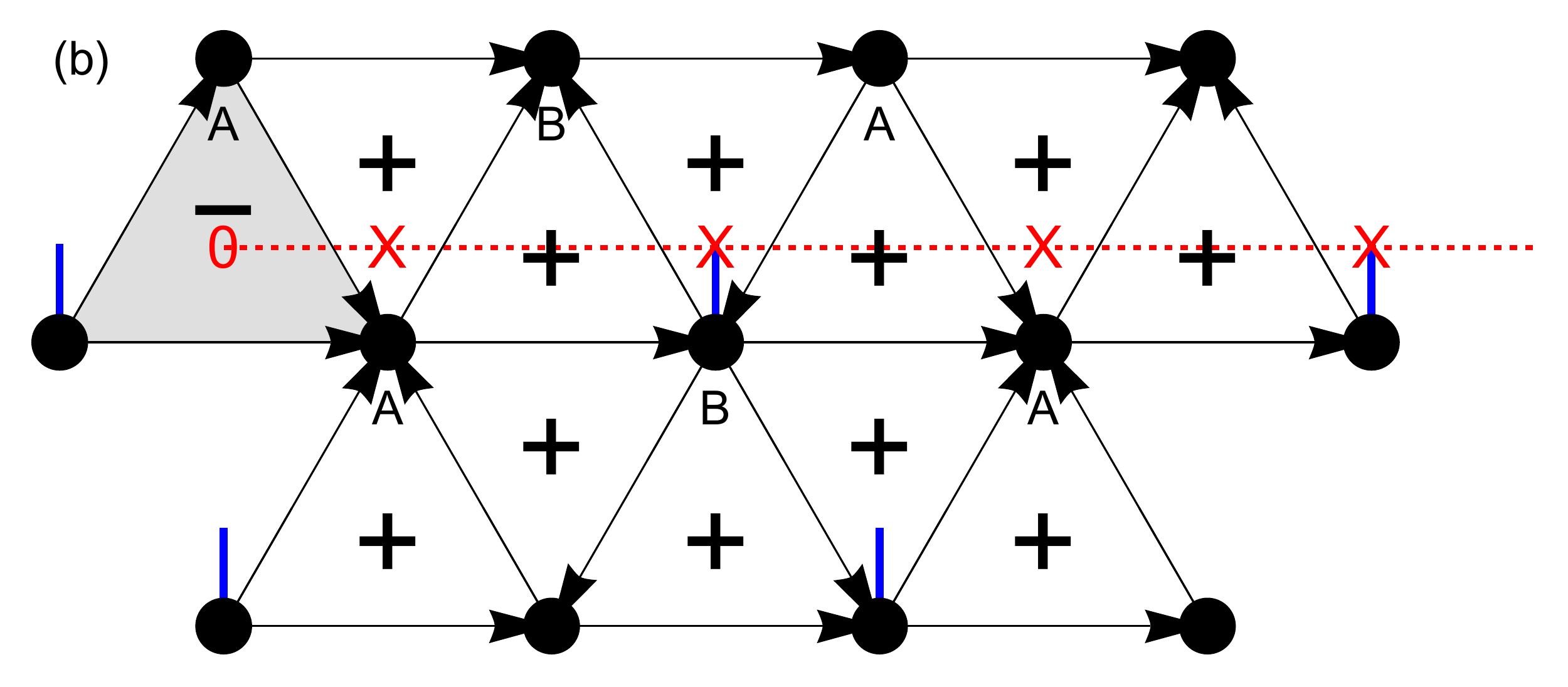}
\caption{\emph{Dual vortex}. (a) Vortex-full sector with a single vortex-free plaquette at position $0$ (shown as a gray-shaded hexagon). Flipped $J$-links along BD for the vortex-full gauge are indicated in blue. $J$-links that need to be further flipped in order to create a single vortex-free plaquette are indicated by a semi-infinite string of red crosses. The unit cell is shown in dashed green. (b) When $J=0$, the triangular sublattices are decoupled. The black triangular sublattice hosts a single ``$-$" triangle in a sea of ``$+$" triangles. This is the ``dual vortex'' (shown as a gray-shaded triangle).}
\label{fig.carlo_vortex}
\end{figure}

A way to create a dual vortex is to start from the vortex-full gauge $u_{jk}^\text{full}$ of Fig.~\ref{fig:vortexfull}(a) and to flip all the $J$ links marked with a red cross in Fig.~\ref{fig.carlo_vortex}(a) along a semi-infinite line. This leads to a gauge, which we call $u_{jk}^\text{dual}$, that creates a single dual vortex (i.e. a vortex-free plaquette isolated in a vortex-full sea), while the other vortex is rejected at infinity. Note that links that are flipped twice (i.e. in blue and with a red cross) are back to their original value in the Kitaev gauge. Still with $J=0$, the corresponding Hamiltonian is now $H=H_{\kappa\text{-bulk}} + H_{\kappa\text{-string}}$.

The triangular lattice is formed by the $\kappa$ links connecting the black (or white) sites. Flipping the $J$-links also flips the $\kappa$ links. In the same way as described above, this creates an anti-vortex in the triangular lattice, at the left end of the red line in Fig.~\ref{fig.carlo_vortex}(b). For the triangular lattice at a quarter-flux quantum per triangle, it means that one triangle host a defect with flux $-\pi/2$ instead of $+\pi/2$ centered around the origin marked `0' in Fig.~\ref{fig.carlo_vortex}(b). To find the state bound to this triangle, we notice that the Hamiltonian in the presence of the dual vortex has two parts $\mathcal{H} = \mathcal{H}^{AB}_{\xi} +\mathcal{H}_{\kappa\text{-string}}$. One $\mathcal{H}^{AB}_{\xi}$, also called the $\kappa$-bulk part, that creates the vortex full triangular lattice, and the other $\mathcal{H}_{\kappa\text{-string}}$ that creates the string of flipped links (both of these terms are proportional to $\kappa$). We also notice that the role of these flipped links is to change the value of trapped bound state wavefunction along the flipped links. To have a more qualitative understanding of these flipped links on the bound state wavefunction, we note that  $u_{ij}$ acts as a nearest-neighbor hopping parameter in Kitaev's honeycomb model. And the hopping parameters between point $i$ and $j$ can be written as the matrix element of the Hamiltonian between the wavefunction at site $i$ and site $j$, $u_{ij} \propto \langle \psi_{i} | \mathcal{H}^{AB}_{\xi} | \psi_{j} \rangle$. Thus, flipping $u_{ij}$ from $+1$ to $-1$ is the same as absorbing this extra minus sign in the wavefunction while keeping the Hamiltonian intact $\langle \psi_{i} | \mathcal{H}^{AB}_{\xi} | -\psi_{j} \rangle$. Thus, the wave functions at the two sites, joined by a flipped link, are in anti-phase. In the continuum limit, one can say that the wave function has a branch cut along the string of flipped links~\cite{Otten19}.

To find the bound-state wavefunction, we first rewrite $\mathcal{H}^{AB}_{\xi}(\mathbf{q})$ of \eqref{brillouinexp} in real space using the substitution $\mathbf{q}\to -i\boldsymbol{\partial}$:
\begin{equation}
    \mathcal{H}^{AB}_{\xi}(\mathbf{r}) = 2\kappa[\sigma_{x}(-i~\mathbf{n}_{3}\cdot \boldsymbol{\partial}) + \xi~ \sigma_{y}(-i~\mathbf{n}_{1}\cdot \boldsymbol{\partial}) + \xi \sigma_{z}].
\end{equation}
Then, we look for a zero energy solution of
\begin{eqnarray}
    \mathcal{H}^{AB}_{\xi}(\mathbf{r})~\psi_\xi = 0 
    \label{mzm_smallj}
\end{eqnarray}
for the two-component spinor $\psi_\xi = \left(\  \begin{array}{cc} 
    \psi^{A}_{\xi}\\
    \psi^{B}_{\xi}
\end{array}\right)$. In order to account for the branch cut, we impose anti-periodic boundary conditions (ABC), that read $\psi_\xi(r,\phi+2\pi)=-\psi_\xi(r,\phi)$ in polar coordinates.

In Appendix~\ref{app:smallJ}, we show that the solutions are:
\begin{eqnarray}
\psi_+ &=& \left(
\begin{array}{c}
\psi^{A}_{+}\\
\psi^{B}_{+}
\end{array}
\right)  = g(r') \left(
\begin{array}{c}
 e^{-i \frac{\phi'}{2}}\\
 i~e^{i \frac{\phi'}{2}}
\end{array}
\right),\notag\\
\psi_- &=&\left(
\begin{array}{c}
\psi^{A}_{-}\\
\psi^{B}_{-}
\end{array}
\right)  = g(r') \left(
\begin{array}{c}
 i~e^{i \frac{\phi'}{2}}\\
 e^{-i\frac{\phi'}{2}}
\end{array}
\right),
\label{carlomzm}
\end{eqnarray}
where $x' = x - \frac{1}{\sqrt{3}}y$,
$y' = \frac{2}{\sqrt{3}}y$, $r' = \sqrt{x'^2 + y'^2}$, $\phi' = \tan^{-1}{(\frac{y'}{x'})}$, $g(r') =\mathcal{N} \frac{1}{\sqrt{r'}} e^{-r'}$, and $\mathcal{N}$ is a normalization constant. These solutions satisfy ABC in $\phi'$: $\psi_\xi(r',\phi'+2\pi)=-\psi_\xi(r',\phi')$. But when $\phi'$ makes a full turn from 0 to $2\pi$, so does $\phi$, so that $\psi_\xi(r,\phi+2\pi)=-\psi_\xi(r,\phi)$, as required.

We can also write down the Hamiltonian which is valid around the $\mathbf{k}_+$ and $\mathbf{k}_-$ points, and it looks like
\begin{eqnarray}
    \mathcal{H}^{AB} = \left( \begin{array}{cc}
         \mathcal{H}^{AB}_{+} & 0  \\
         0 & \sigma_x\mathcal{H}^{AB}_{-}\sigma_x
    \end{array} \right)
\end{eqnarray}
in a basis $\left(\begin{array}{cccc} c_{\mathbf{k}_+, A} & c_{\mathbf{k}_+, B} & c_{-\mathbf{k}_+, B} & c_{-\mathbf{k}_+, A} \end{array}\right)^T$ (the inversion $A,B\to B,A$ in the valley $\xi=-$ is the reason for the appearance of the $\sigma_x$'s in the above equation).
The MZM localised at a single plaquette of the black triangular sublattice can thus be written as the solution of 
\begin{eqnarray}
    \mathcal{H}^{AB}(\mathbf{r})~\psi_b = 0,  
\end{eqnarray}
where $\psi_b$ is the normalized four-component spinor
\begin{eqnarray}
\psi_b (\mathbf{r}) 
             & = &
             g(r') \left( \begin{array}{cccc}  e^{i(-\frac{\phi'}{2} + \mathbf{k}_+\cdot\mathbf{r} - \frac{\pi}{4}+\theta)} \\
             e^{i(\frac{\phi'}{2} + \mathbf{k}_+\cdot\mathbf{r}+\frac{\pi}{4}+\theta)}\\
             e^{-i(\frac{\phi'}{2} + \mathbf{k}_+\cdot\mathbf{r} + \frac{\pi}{4}+\theta)}\\
             e^{-i(-\frac{\phi'}{2} + \mathbf{k}_+\cdot\mathbf{r}-\frac{\pi}{4}+\theta)}
             \end{array}\right)
              \label{MZMb}
\end{eqnarray}
written in the basis $\left(\begin{array}{cccc} c_{\mathbf{k}_+, A} & c_{\mathbf{k}_+, B} & c_{-\mathbf{k}_+, B}^{} & c_{-\mathbf{k}_+, A} \end{array}\right)^T$, with 
\begin{equation}
g(r') = \frac{e^{-r'}}{\sqrt{2\sqrt{3}\pi r'}}.
\end{equation}
We allowed for a possible relative phase $\theta$ between the two valleys at $\mathbf{k}_+$ and $-\mathbf{k}_+$. 
The wavefunction on sublattice $A$, for example, is 
\begin{equation} \label{eq:lwf}
\psi_b^A(\mathbf{r})=g(r')2\cos(\mathbf{k}_+\cdot\mathbf{r} -\frac{\phi'}{2}- \frac{\pi}{4}+\theta).
\end{equation}
Mirror symmetry with respect to the line passing through the dual vortex and perpendicular to $\mathbf{k}_+$ imposes that $\psi_b^A(x,-y)=\pm \psi_b^A(x,y)$, so that $\theta=\pi/4$ modulo $\pi/2$. 

From \eqref{MZMb}, we can construct the zero mode operator as
\begin{align}
    \gamma  &= \int d^2 \mathbf{r}~ g(r') \times \\
      &[ e^{i(-\frac{\phi'}{2} + \mathbf{k}_+\cdot\mathbf{r} - \frac{\pi}{4}+\theta)}~c_{\mathbf{k}_+, A} + e^{-i(-\frac{\phi'}{2} + \mathbf{k}_+\cdot\mathbf{r} - \frac{\pi}{4}+\theta)}~c_{\mathbf{k}_+, A}^\dagger \notag \\
     &+  e^{i(\frac{\phi'}{2} + \mathbf{k}_+\cdot\mathbf{r}+\frac{\pi}{4}+\theta)}~c_{\mathbf{k}_+, B} + e^{-i(\frac{\phi'}{2} + \mathbf{k}_+\cdot\mathbf{r}+\frac{\pi}{4}+\theta)}~c_{\mathbf{k}_+, B}^\dagger  ], \notag \label{mzmoperator}
\end{align}
which is self-adjoint, as required for a Majorana mode. 

The above analysis was done for the black triangular lattice formed by $A$ and $B$ sites. For the white triangular lattice, formed by sites $C$ and $D$, one obtains the following effective Hamiltonian
\begin{eqnarray}
\mathcal{H}^{CD}_{\xi} (\mathbf{q})
=2\kappa\left[-\mathbf{q}\cdot\mathbf{n_{3}}\sigma_x + \xi \mathbf{q}\cdot\mathbf{n_{1}}\sigma_y - \xi \sigma_z\right].
  \label{wtriangle}
\end{eqnarray}
Note that now, the chirality $\mathcal{C}_\xi=-\xi$ and $\text{sign}(m_\xi) = -\xi$ so that the Chern number is also $\nu=-1$.
Following the previous method, we can obtain the expression for the wavefunction of MZM trapped in the vortex of a white triangular lattice as
\begin{eqnarray}
\psi_w (\mathbf{r}) & = g(r') \left( \begin{array}{cccc}  e^{i(\frac{\phi'}{2} + \mathbf{k}_+\cdot\mathbf{r} - \frac{\pi}{4}+\theta)} \\
             e^{i(-\frac{\phi'}{2} + \mathbf{k}_+\cdot\mathbf{r}+\frac{\pi}{4}+\theta)}\\
             e^{-i(-\frac{\phi'}{2} + \mathbf{k}_+\cdot\mathbf{r} + \frac{\pi}{4}+\theta)}\\
             e^{-i(\frac{\phi'}{2} + \mathbf{k}_+\cdot\mathbf{r}-\frac{\pi}{4}+\theta)}
             \end{array}\right)
             \label{MZMw}
\end{eqnarray}
written in the basis $\left(\begin{array}{cccc} c_{\mathbf{k}_+, C} & c_{\mathbf{k}_+, D} & c_{-\mathbf{k}_+, D} & c_{-\mathbf{k}_+, C} \end{array}\right)^T$ and with the same $\theta=\pi/4$ modulo $\pi/2$. 


When comparing the analytical wavefunction obtained in the continuum model with the numerical wavefunction obtained by exact diagonalization on the lattice, we need to account for the fact that the different sites in the unit cell do not exactly have the same position. For example, on the black triangular sublattice
\begin{eqnarray}
\psi_b(\mathbf{r})             & \simeq  
             \left( \begin{array}{cccc}  g(r_A')  e^{i(-\frac{\phi_A'}{2} + \mathbf{k}_+\cdot\mathbf{r} - \frac{\pi}{4}+\theta)} \\
             g(r_B') e^{i(\frac{\phi_B'}{2} + \mathbf{k}_+\cdot\mathbf{r}+\frac{\pi}{4}+\theta)}\\
             g(r_B') e^{-i(\frac{\phi_B'}{2} + \mathbf{k}_+\cdot\mathbf{r} + \frac{\pi}{4}+\theta)}\\
             g(r_A') e^{-i(-\frac{\phi_A'}{2} + \mathbf{k}_+\cdot\mathbf{r}-\frac{\pi}{4}+\theta)}
             \end{array}\right),
\end{eqnarray}
with $\mathbf{r}_A = \mathbf{r}+\boldsymbol{\delta}_A$, where $\boldsymbol{r}$ is the position of the unit cell (Bravais lattice position) and $\boldsymbol{\delta}_A$ is the position of the $A$ site within the unit cell. Then the relation between unprimed $\mathbf{r}_A=(x_A,y_A)$ and primed $\mathbf{r}_A'=(x_A',y_A')$ coordinates is given by Eq.~\eqref{primed}. And finally, the primed position can be given in polar coordinates according to $x_A'+i y_A' = r_A' e^{i \phi_A'}$.

\subsubsection{Energy splitting}\label{sec:energysplitting}
Now, we turn on the coupling $J$ so that the total Hamiltonian is given by Eq.~\eqref{eq:totham}. This introduces both a $J$-bulk term (corresponding to the vortex-full pattern) and a $J$-string term (corresponding to the semi-infinite string of flipped links). 

\paragraph{$J$-string}
We start with the Hamiltonian generating the string: 
\begin{eqnarray}
    H_{J\text{-str.}} & =& - 2 iJ \sum_{n=0}^\infty (c_{n, A}c_{n, C}-c_{n, C}c_{n, A} \notag \\ 
    &+& c_{n, D}c_{n, B} - c_{n, B}c_{n, D}), \label{Jstringc}
\end{eqnarray} 
where $n$ labels unit cells along the positive $x$ axis starting from the position of the dual vortex. To evaluate this sum, we choose a unit cell oriented along the $x$-axis that contains the four sites $A$, $B$, $C$, $D$ [see Fig.~\ref{fig.carlo_vortex}(a)] and the unit cell position is $\mathbf{r}_n=(2n,0)$. 

This Hamiltonian needs to be written in the continuum model. For that purpose, one uses a transformation that relates lattice to continuum-limit operators~\cite{Otten19,CastroNeto09}:
\begin{eqnarray}
    c_{n,\alpha}\simeq  \sqrt{\mathcal{A}} \left[e^{i\mathbf{k}_+\cdot \mathbf{r}_{n}}  c_{\alpha}(\mathbf{r}_{n}) +  e^{-i\mathbf{k}_+\cdot \mathbf{r}_{n}}  c_{\alpha}^{\dagger}(\mathbf{r}_{n})\right]. \label{c_to_l}
\end{eqnarray}
Here $c_{n,\alpha}$ is a lattice operator that annihilates a particle on sublattice $\alpha= A, B, C$ or $D$ in the unit cell at position $\mathbf{r}_{n}$, whereas $c_{\alpha}(\mathbf{r}_{n})$ is a continuum operator that varies slowly over the unit cell (similar to an envelope function) and that describes the contribution close to the momentum $\mathbf{k}_+$, and  $\mathcal{A}=\sqrt{3}$ is the area of the unit cell (corresponding to two hexagons of the honeycomb).
We can arrive at this form by starting from the Fourier transform of $c_{n,\alpha}$, selecting the low-energy contribution in the vicinity of the Dirac points $\mathbf{k}_\pm$ and remembering that $c_{\mathbf{k},\alpha}^{\dagger} = c_{-\mathbf{k},\alpha}$. 
Substituting this into the equation~(\ref{Jstringc}) results in the continuum version of $H_{\text{J-str.}}$, written in the basis 
$\mathbf{C}(\mathbf{r}) = [c_{A}(\mathbf{r}_{A}), c_{B}(\mathbf{r}_{B}), c_{B}^{\dagger}(\mathbf{r}_{B}), c_{A}^{\dagger}(\mathbf{r}_{A}),\\
c_{C}(\mathbf{r}_{C}), c_{D}(\mathbf{r}_{D}), c_{D}^{\dagger}(\mathbf{r}_{D}),c_{C}^{\dagger}(\mathbf{r}_{C})]^{T}$:
\begin{eqnarray}\label{Jstringcon}
    H_{J\text{-str.}} 
    &=& \int d^2\mathbf{r} \, \mathbf{C}^{\dagger}(\mathbf{r}) \mathcal{H}_{J\text{-str.}} \mathbf{C}(\mathbf{r}),
\end{eqnarray}
where
\begin{eqnarray}
\mathcal{H}_{J\text{-str.}} & =& -\sqrt{3} i J \Theta(x) \delta(y)  \nn \\
&\times & \begin{pmatrix}
0 &  0 & 0 & 0 & 1 & 0 & 0 & 0 \\
0  & 0 & 0 & 0 & 0 &-1 & 0 & 0 \\
0  & 0 & 0 & 0 & 0 & 0 &-1 & 0 \\
0  & 0 & 0 & 0 & 0 & 0 & 0 & 1 \\
1  & 0 & 0 & 0 & 0 & 0 & 0 & 0 \\
0  &-1 & 0 & 0 & 0 & 0 & 0 & 0 \\
0  & 0 &-1 & 0 & 0 & 0 & 0 & 0 \\
0  & 0 & 0 & 1 & 0 & 0 & 0 & 0
\end{pmatrix}.
\label{Jstring}
\end{eqnarray}
{
$H_{J\text{-str.}}$ couples the two MZMs, one trapped at the vortex in the white triangular lattice and the other trapped at the vortex of the black triangular lattice. 

We write these two MZM wavefunctions as eight-component spinors:
\begin{eqnarray}
\chi_{1} & = \left( \begin{array}{cc} \psi_b\\
    0
    \end{array}\right), \,
        \chi_{2} & = \left( \begin{array}{cc} 0 \\
    \psi_w
    \end{array}\right) \label{mzmwb}
\end{eqnarray}
We find that the matrix element due to the $J$-string is:
\begin{align}
    \langle \chi_1| H_\text{$J$-str.} |\chi_2\rangle  
    &\simeq 2iJ\sqrt{3} \int_{0}^{\infty} \!\!\!\!\!\! dx \, [\textsl{g}(x+\frac{2}{3},-\frac{1}{3})\textsl{g}(x+\frac{1}{3},\frac{1}{3})\nn\\
    &-\textsl{g}(x+\frac{5}{3},-\frac{1}{3})\textsl{g}(x+\frac{4}{3},\frac{1}{3})]\nn \\
    &\simeq  i\, 0.052\, J,
\end{align}
where $\textsl{g}(x',y')\equiv g(r')$ with $r'=\sqrt{x'^2+y'^2}$. 

\paragraph{$J$-bulk}
The continuum bulk Hamiltonian $H_{J\text{-bulk}}$ can be obtained in two equivalent ways. In the first approach, we start from the lattice Hamiltonian $H_{J}$~(\ref{kitaevhev}), and express the lattice Majorana operators in terms of slowly varying envelope fields via Eq. (\ref{c_to_l}), retaining only non-oscillating contributions.
Alternatively, we can start with the Bloch Hamiltonian of the vortex-full sector~(\ref{A:kitaevh}) with $J=0$, expand to linear order in $\mathbf{q}$ around the Dirac points as $\mathbf{k} = \pm \mathbf{k}_{+} + \mathbf{q}$, i.e. take the long-wavelength limit. Upon transforming to real space via the standard correspondence $\mathbf{q} \rightarrow -i\boldsymbol{\partial}$, this yields a continuum Majorana Hamiltonian valid close to Dirac point $\pm \mathbf{k}_{+}$. Importantly, both valleys must be retained; the resulting fermion doubling directly reflects the operator content of Eq. (\ref{c_to_l}) and the Majorana reality condition.
Both approaches lead to the same real-space continuum Hamiltonian written in the combined basis 
$\mathbf{C}(\mathbf{r}) = [c_{A}(\mathbf{r}_{A}), c_{B}(\mathbf{r}_{B}), c_{B}^{\dagger}(\mathbf{r}_{B}), c_{A}^{\dagger}(\mathbf{r}_{A}),\\
c_{C}(\mathbf{r}_{C}), c_{D}(\mathbf{r}_{D}), c_{D}^{\dagger}(\mathbf{r}_{D}),c_{C}^{\dagger}(\mathbf{r}_{C})]^{T}$.
Keeping only the non–oscillating contributions (i.e. coupling $\mathbf{k}_+$ to $\mathbf{k}_+$ but not to $-\mathbf{k}_+$), we obtain
\begin{equation}
H_{J\text{-bulk}} =
\int d^2 \mathbf{r}\;
C^{\dagger}(\mathbf r)\,
\mathcal{H}_{J\text{-bulk}}(\mathbf r)\,
C(\mathbf r).
\label{eq:HJ-cont}
\end{equation}
For exact form of the Hamiltonian, refer to Appendix~\ref{app:smalljham}. One can then numerically compute the integral to find that the $J$-bulk matrix element is:
\begin{eqnarray}
    \langle \chi_1| \mathcal{H}_\text{$J$-bulk} |\chi_2\rangle \simeq -(0.616i+0.066)\, J.
\end{eqnarray}
For a calculation done directly on the lattice, and not in a continuum approximation, we would expect a purely imaginary matrix element.

\paragraph{Total splitting}
The total half-splitting is therefore
\begin{equation}
 \epsilon = |\langle \chi_1| H_\text{$J$-str.}+H_\text{$J$-bulk} |\chi_2\rangle|\simeq   0.57~J .
\end{equation}
This should be compared with the numerical result $0.3933\, J$, obtained through exact diagonalization (see Fig.~\ref{fig:EpsilonVsJ_bulkVSstring} and sec.~\ref{sec:fnsmallJ} below). The order of magnitude is correct but the number obtained from our effective model is too large by 45\%. This is probably due to the poor continuum approximation we took: we approximated rather narrow bands separated by a large band gap (and carrying finite Chern numbers) by two gapped Dirac fermions, that actually correspond to saddle points of the dispersion relation (see $\mathbf{k}_\pm$ in Fig.~\ref{fig.BZisoenergy}). 
\begin{figure}[h]
\includegraphics[width=0.9\columnwidth]{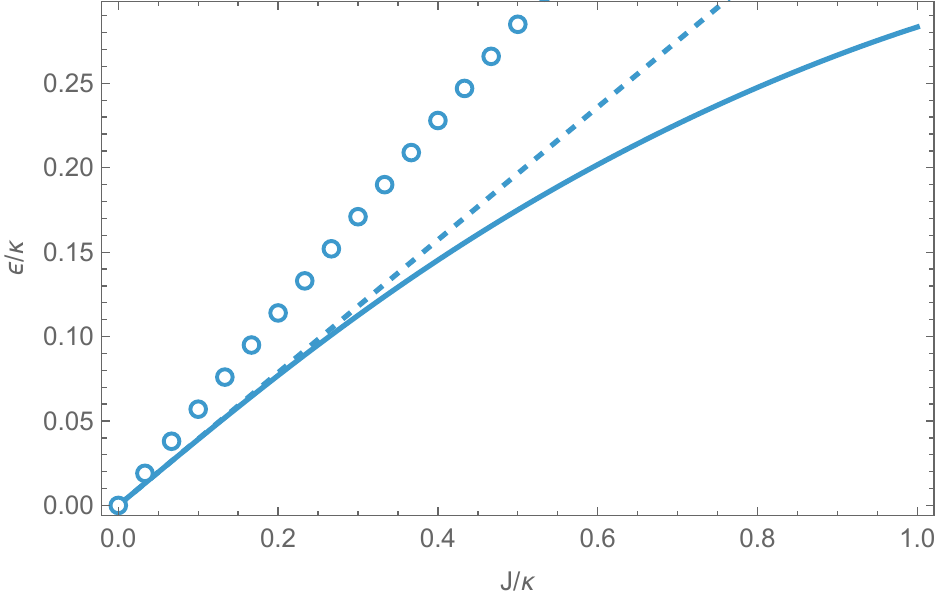}
\caption{\label{fig:EpsilonVsJ_bulkVSstring} Full numerics (continuous lines, with $31^2$ plaquettes on a torus), numerical perturbation theory (dashed lines, with $31^2$ plaquettes on a torus) and analytical perturbation theory (open symbols) for the half-splitting $\epsilon$ as a function of $J$ (both in units of $\kappa$) in the small $J$ limit.}
\end{figure}

\subsection{Numerical perturbation theory}
\label{sec:nptsmallJ}
In order to validate the scenario we have just outlined, we perform perturbation theory numerically directly on the lattice instead of doing it in a continuum approximation. We write the total Hamiltonian $H=H_\kappa + H_J$, where $H_\kappa$ is the zero-order term and $H_J$ is treated as a perturbation.

We first consider the vortex-full problem at $J=0$ on a torus (i.e. with periodic boundary conditions) so that the two triangular sublattices are decoupled. We take an \textit{odd} number of hexagonal plaquettes ($31^2=961$ to be concrete) and start from the Kitaev gauge that corresponds to a vortex-free situation. Next, by flipping links we can introduce pairs of vortices. As the total number of plaquettes is odd, we can never reach a situation in which every plaquette would contain a vortex. At most, we can have 960 plaquettes containing a  vortex and 1 vortex-free plaquette. This allows us to study a \textit{single} dual vortex on a torus.

We then numerically diagonalize  $H_\kappa$ and find two zero modes. The corresponding wavefunctions $\chi_1$ and $\chi_2$ are real and belong each to a different triangular sublattice. We exactly have $H_\kappa|\chi_1\rangle=0$ and $H_\kappa|\chi_2\rangle=0$, and $\langle \chi_1|\chi_2\rangle = 0$. They are plotted in Figure~\ref{fig:MZMwfJ0}. 
\begin{figure}[h]
\includegraphics[width=0.9\columnwidth]{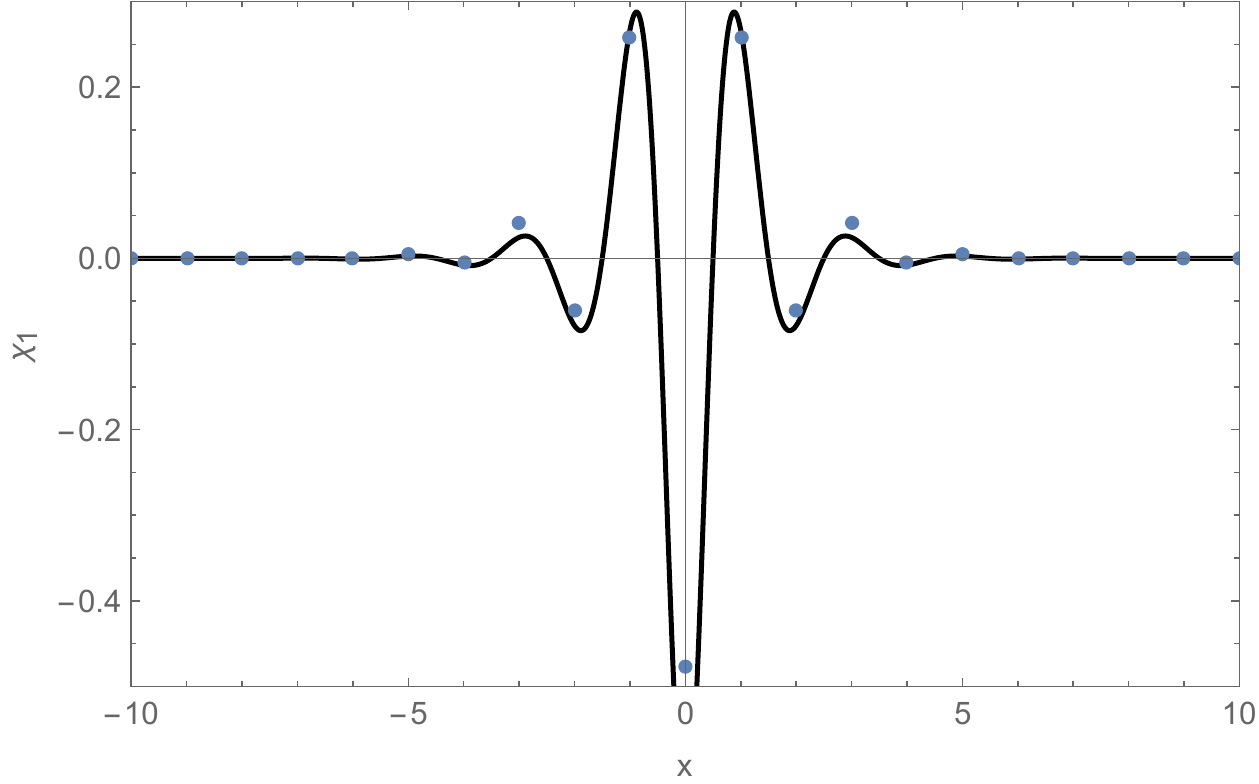}
\includegraphics[width=0.9\columnwidth]{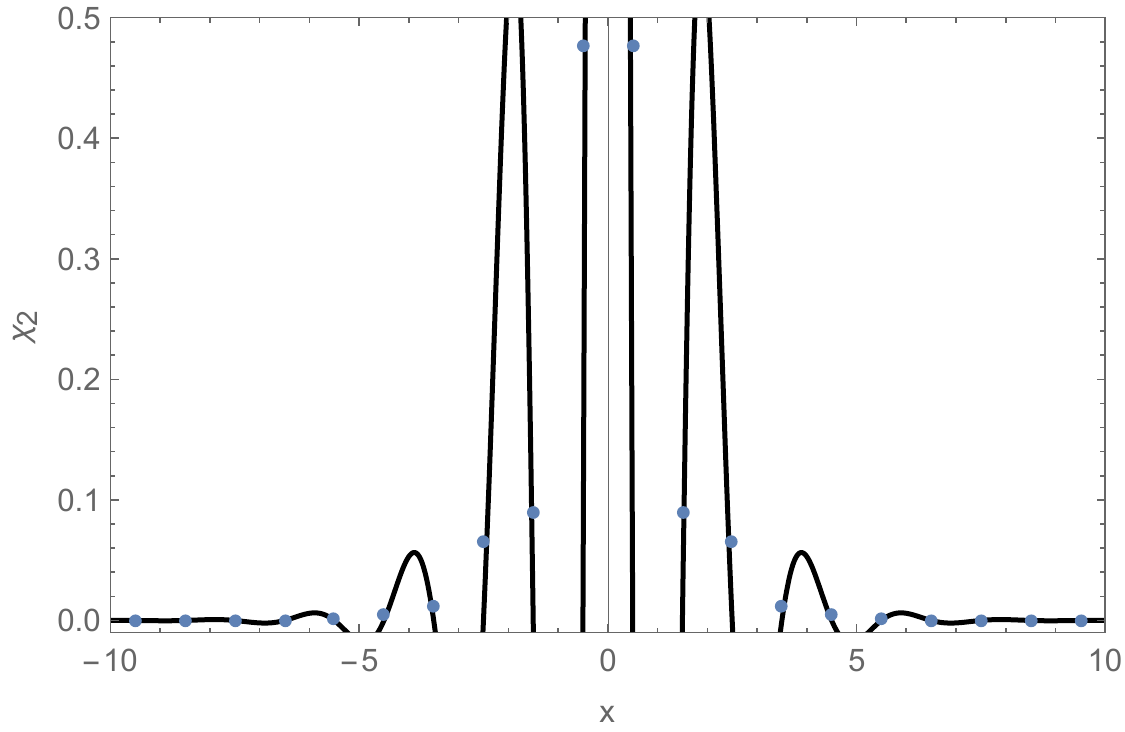}
\caption{\label{fig:MZMwfJ0} Numerically-obtained MZM wavefunctions $\chi_1(x,y)$ and $\chi_2(x,y)$ at fixed $y =1/(2\sqrt{3})$ (just above the dual vortex) for $J=0$ and $31^2$ plaquettes (blue dots). The black line is $g(r)\cos(k_+\, x+\varphi)$ with $k_+=\pi$ and $\varphi=\pi$ (up) or $0$ (down).}
\end{figure}

As compared to the analytical wavefunction [see Eq~\eqref{eq:lwf}],  
the numerical one has the expected $g(r)\propto e^{-r}/\sqrt{r}$ envelope and the lattice scale oscillations, but, as there is a single dual vortex, there is no branch cut. We will therefore compare with the analytical form
\begin{equation} \label{eq:lwfsmallJ}
\psi(\mathbf{r}) = g(r) \cos(\mathbf{k}_+ \cdot \mathbf{r} +\varphi)
\end{equation}
without the branch cut term in $-\phi/2$ [compare with Eq.~\eqref{eq:lwf}], where $\varphi$ is a phaseshift corresponding roughly to $-\pi/4+\theta$. Inversion symmetry imposes that $\psi(-\mathbf{r})=\pm \psi(\mathbf{r})$ so that $\varphi=0,\pm \pi/2$ or $\pi$. In Figure~\ref{fig:MZMwfJ0}, we compare the numerically obtained wavefunctions with \eqref{eq:lwfsmallJ} in which $\mathbf{k}_+$ and $\varphi$ are treated as fitting parameters. The agreement is quite good. We cannot expect a perfect match as the wave functions are gauge-dependent and the analytical and numerical calculations are obtained in two very different gauges. Also the analytical wave functions were obtained in a long wavelength approximation that captures the Chern number but not the correct dispersion relation (cf. the discussion about the saddle point versus gapped Dirac cone). 

The $J=0$ situation when the two triangular sublattices are decoupled deserves a special discussion. Let us for a moment forget about the Majorana problem and consider a tight-binding problem of an electron hopping on a triangular lattice in the presence of a $\pm \pi/2$ flux in every triangle. It can be written in a gauge such that every hopping term is purely imaginary (one just needs to choose the Peierls phase to be $\pm \pi/2$ for each hopping amplitude). The tight-binding Hamiltonian is then purely imaginary (and antisymmetric) so that the energy spectrum has the $E \to -E$ symmetry and the corresponding eigenstates are $\psi$ and $\psi^*$. If the system is placed on a torus and contains an \textit{odd} number of sites (say $31^2=961$), one cannot find a gauge such that every triangle has a $+\pi/2$ flux. The reason is that the total magnetic flux across the torus must be an integer multiple of $2\pi$ because of the Dirac magnetic monopole quantization condition (see e.g. Appendix A in~\cite{Fuchs16}). Therefore, the best we can do is to have 960 triangles with $+\pi/2$ flux and 1 triangle with flux $-\pi/2$.
Because there is an odd number of eigenvalues and the $E\to -E$ symmetry, there must be at least one zero eigenvalue. This corresponds to a zero-energy state bound to the defect triangle containing a $-\pi/2$ flux. One should resist the temptation of interpreting this Hamiltonian as a Majorana Hamiltonian with a \textit{single} MZM: as it has an odd dimension, this is not possible. In the Kitaev honeycomb model at $J=0$, we actually have two triangular sublattices and therefore exactly two decoupled MZM.

Next, we turn on a small finite $J$ and check that $\langle\chi_1| H_J|\chi_1\rangle=0$ (and similarly for $\chi_2$) because the $J$-term couples the two sublattices but does not couple $\chi_1$ to $\chi_1$ or $\chi_2$ to $\chi_2$. Then, we numerically compute the matrix element that couples them to find
\begin{equation}
\langle\chi_1| H_J|\chi_2\rangle\simeq  
i\, 0.3933\, J
\end{equation}
so that the half-splitting in numerical perturbation theory is $\epsilon \simeq   0.3933 J$. 
See Figure~\ref{fig:EpsilonVsJ_bulkVSstring} for a comparison with analytical perturbation theory.

\subsection{Full numerics}\label{sec:fnsmallJ}
We have also performed a fully numerical (non-perturbative) calculation of the splitting as a function of $J$. For this, we take the total Hamiltonian in the presence of a single dual vortex  on a torus with $31^2$ plaquettes.
Diagonalizing it, we find the half-splitting $\epsilon$ versus $J$ plotted in Fig.~\ref{fig:EpsilonVsJ_bulkVSstring} as a continuous line. In the small $J$ limit, $\epsilon$ is linear with a slope $\epsilon/J \simeq 0.3933$ that agrees with the numerical perturbation theory.

\subsection{Conclusion}
At $J=0$, there are two decoupled triangular sublattices with $\pi/2$ flux per triangle. For each sublattice, the $\kappa$ Hamiltonian produces a particle-hole symmetric band structure made of two narrow bands separated by a large bulk gap and with a Chern number of $-1$ (equivalent to a $p_x+ip_y$ superconductor). One can nevertheless identify two massive Dirac fermions, that actually correspond to saddle points of the dispersion relation. Introducing a dual vortex via a semi-infinite $\kappa$-string results in the appearance of one MZM bound to the dual vortex in each sublattice. Next, turning on a small finite $J$ couples the two triangular sublattices and therefore the two Majorana modes. The main contribution to the energy splitting comes from the $J$-string and only 10\% is due to the bulk $J$ term.

\section{Small $\kappa$ limit}
\label{sec:smallkappa}
In this section, we study the $\kappa\ll J$ limit. The Hamiltonian is the sum of a bulk and a string term. It will turn out that the two MZMs are localized in different regions of the Brillouin zone. Then, the string term can further be split according to whether it oscillates or not in real space, i.e. whether or not it can couple reciprocal space points that are distant. We therefore split the total Hamiltonian~\eqref{eq:totham} in three terms according to:
\beqn
H =  H_{\text{bulk}} + H_{\text{string}}^\text{nosc} + H_{\text{string}}^\text{osc}.
\label{eq:3hams}
\eeqn

In section~\ref{sec:anasmallkappa}, we do analytical perturbation theory in a continuum approximation: we first consider the bulk term $H_{\text{bulk}}$ with $0<\kappa\ll J$ that produces four gapped Dirac cones (section \ref{sec:gappedDirac}), then add the non-oscillating string $H_{\text{string}}^\text{nosc}$ that creates two uncoupled MZMs (section \ref{sec:dualvortex2}), and eventually add the oscillating string $H_{\text{string}}^\text{osc}$ that couples the Majorana modes (section \ref{sec:energysplitting2}).  We then check our analytical results by performing numerical perturbation theory directly on the lattice (section \ref{sec:nptsmallkappa}) and a fully numerical calculation on a finite-size system (section \ref{sec:fnsmallkappa}).

\subsection{Analytic perturbation theory in the continuum}
\label{sec:anasmallkappa}
\subsubsection{Gapped Dirac cones}\label{sec:gappedDirac}
\begin{figure}[!h]
\includegraphics[width=\linewidth]{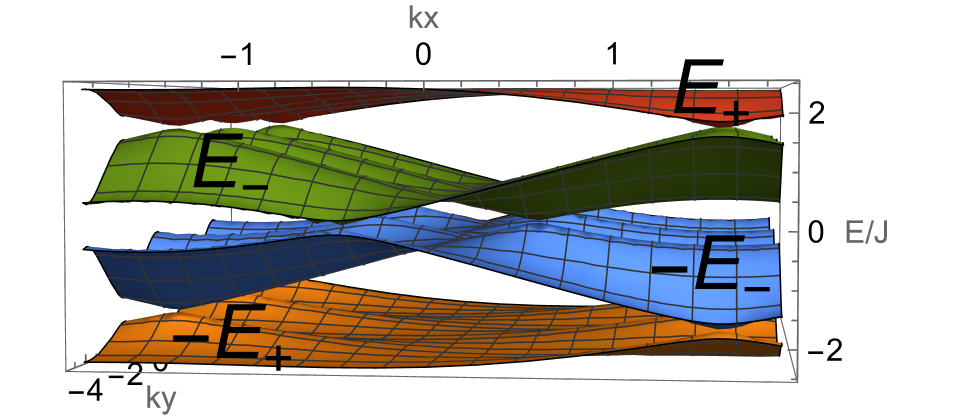}
\includegraphics[width=\linewidth]{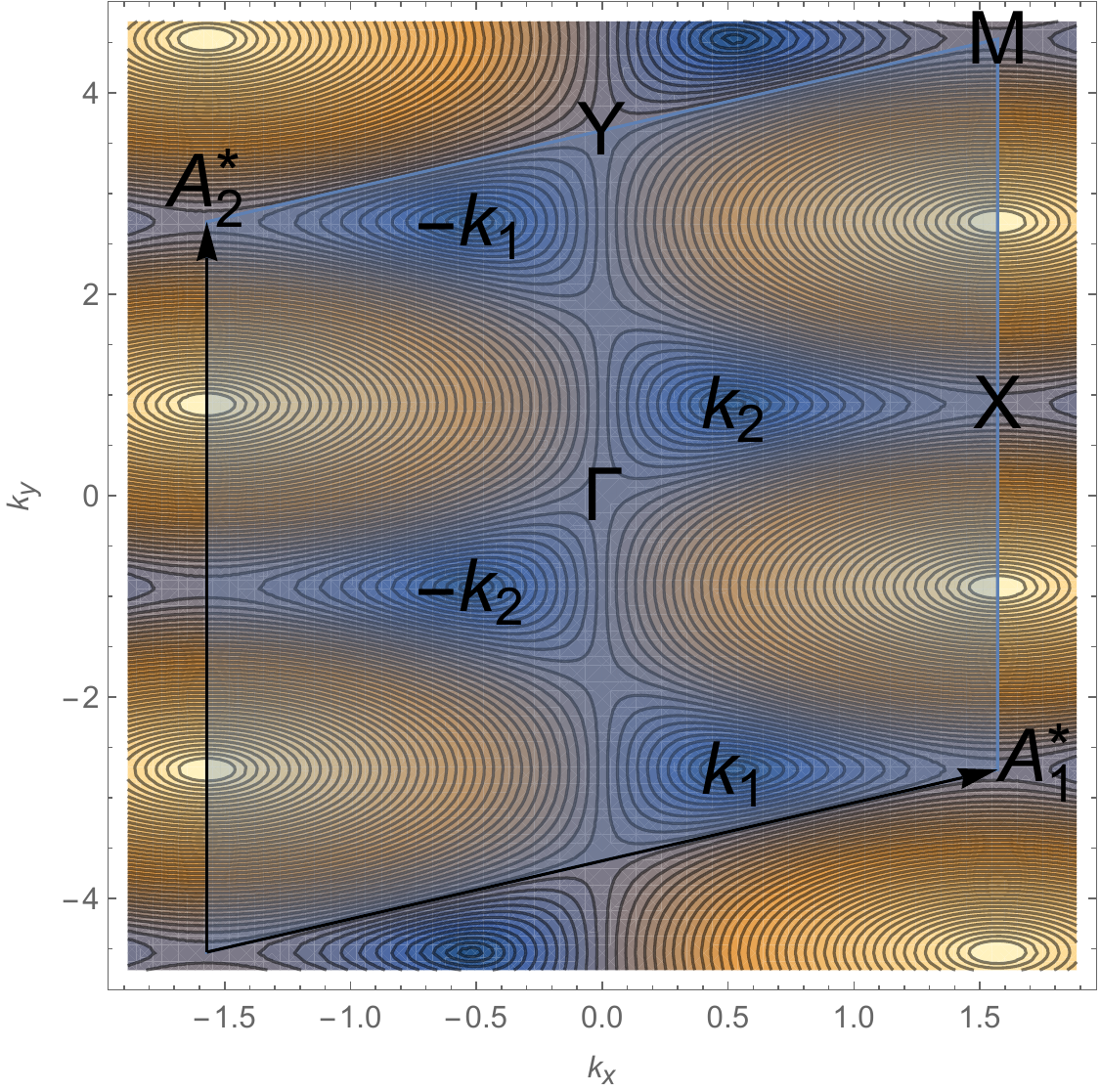}
\caption{(Top) Energy bands $\pm E_-(\mathbf{k})$ and $\pm E_+(\mathbf{k})$ (in units of $J$), see Eq.~\eqref{eq:dispvf}. (Bottom) Isoenergy lines of $E_-(\mathbf{k})$ in reciprocal space (at $\kappa=0$). Also indicated are the Brillouin zone, the reciprocal lattice vectors $\mathbf{A}_1^*$ and $\mathbf{A}_2^*$, the high symmetry points $\Gamma$, $X$, $Y$ and $M$ and the position of four Dirac points $\pm \mathbf{k}_1$ and $\pm \mathbf{k}_2$.}
\label{fig.BZisoenergy_kappa0}
\end{figure}
We start from the Kitaev Hamiltonian in the vortex-full sector, see Appendix~\ref{app:vortexfulldr}. When $\kappa\ll J$, there are four energy bands, among which, two $\pm E_-$ are close to zero energy and the other two $\pm E_+$ are at higher energy (see Fig.~\ref{fig.BZisoenergy_kappa0}). In this section we use energy units such that $J=1$. When $\kappa=0$, there are four Dirac points at zero energy (i.e. points where $E_-(\mathbf{k})$ vanishes) at
$\pm \mathbf{k}_{1} = \pm (\frac{1}{6}\mathbf{A}_{1}^* - \frac{5}{12} \mathbf{A}_{2}^*) = \pm (\frac{\pi}{6},-\frac{\pi \sqrt{3}}{2})$ and
$\pm \mathbf{k}_{2} = \pm (\frac{1}{6}\mathbf{A}_{1}^* + \frac{1}{12} \mathbf{A}_{2}^*) = \pm (\frac{\pi}{6},\frac{\pi}{2\sqrt{3}})$, 
A finite $\kappa$ opens up a gap at these points~\cite{Lahtinen10}. Since we are interested in the low-energy physics, we construct an effective theory by projecting onto the two bands closest to zero energy and expanding around the Dirac points. The low energy excitations are localized in momentum space near these four points. This procedure involves:
\begin{itemize}
    \item Projection onto the low energy subspace of each Dirac point, e.g. $+\mathbf{k}_1$.
    \item Expansion of the Bloch Hamiltonian to linear order in $\mathbf{q} = \mathbf{k} - \mathbf{k}_{1}$.
    \item A continuum limit obtained by the substitution $\mathbf{q} \rightarrow -i \mathbf{\partial}$.
\end{itemize}
A detailed derivation is provided in  Appendix~\ref{app:smallkappa}. In the end, since these regions (close to each Dirac point) are well separated in momentum space, the corresponding degrees of freedom are independent at low energies and can be combined into a single spinor. This leads to the extended basis
\begin{eqnarray}
 \mathbf{L}(\mathbf{r}) & = (L_{1}^{\mathbf{k}_{1}} (\mathbf{r}), L_{2}^{\mathbf{k}_{1}} (\mathbf{r}), L_{2}^{\mathbf{k}_{1}\dagger} (\mathbf{r}), L_{1}^{\mathbf{k}_{1}\dagger} (\mathbf{r}), \notag \\
 & L_{1}^{\mathbf{k}_{2}} (\mathbf{r}), L_{2}^{\mathbf{k}_{2}} (\mathbf{r}), L_{2}^{\mathbf{k}_{2}\dagger} (\mathbf{r}), L_{1}^{\mathbf{k}_{2}\dagger} (\mathbf{r}))^{T},
 \label{l_basis}
\end{eqnarray}
made of two bands $i=1,2$ for each of the 4 Dirac points $\pm \mathbf{k}_1$ and $\pm \mathbf{k}_2$, the effective Hamiltonian has the form
\begin{eqnarray}
    H(\mathbf{r}) = \int d^2 \mathbf{r}  \; \mathbf{L}^T (\mathbf{r}) \mathcal{\bar{H}}(\mathbf{r}) \mathbf{L} (\mathbf{r}); \label{kitaevhcl}
\end{eqnarray}
where $\mathcal{\bar{H}}(\mathbf{r})$ has a block diagonal structure
\begin{eqnarray}
    \mathcal{\bar{H}}(\mathbf{r}) = \left(\  \begin{array}{cc} \mathcal{\bar{H}}_{\mathbf{k}_1}(\mathbf{r}) & 0\\
    0 & \mathcal{\bar{H}}_{\mathbf{k}_2} (\mathbf{r})\end{array}\right). \label{hfull}
\end{eqnarray}
Each block $\mathcal{\bar{H}}_{\mathbf{k}_1,2}(\mathbf{r})$ itself consists of contribution from two valleys $\pm \mathbf{k}_{1,2}$ and takes the form
\begin{eqnarray}
    \mathcal{\bar{H}}_{\mathbf{k}_{1(2)}}(\mathbf{r}) = \left(\  \begin{array}{cc} \bar{\mathcal{H}}_{+\mathbf{k}_{1(2)}}(\mathbf{r}) & 0\\
    0 & - \, \sigma_{z} \bar{\mathcal{H}}^{T}_{-\mathbf{k}_{1(2)}}  \sigma_{z} \end{array}\right).\label{dirack1}
\end{eqnarray}
The matrices $\bar{\mathcal{H}}_{\xi\mathbf{k}_{1(2)}}(\mathbf{r})$, with $\xi = \pm 1$, are continuum Dirac Hamiltonians obtained after projection onto the low energy subspace and expansion around the corresponding Dirac points. Explicitly, they read
\begin{eqnarray}
  \bar{\mathcal{H}}_{\xi\mathbf{k}_{1}}(\mathbf{r}) = 
  \scalebox{0.83}{$
  \left(\  \begin{array}{cc} 
    -\xi \sqrt{3} \kappa &  -\frac{i e^{\xi i \frac{11\pi}{12}}}{\sqrt{2}}  (\partial_{x} - \xi i \partial_{y}) \\
    -\frac{i e^{-i \xi \frac{11\pi}{12}}}{\sqrt{2}}  (\partial_{x} + \xi i \partial_{y}) & \xi\sqrt{3} \kappa 
    \end{array}\right)
    $}
    \label{dhr1}
\end{eqnarray}
and 
\begin{eqnarray}
  \bar{\mathcal{H}}_{\xi\mathbf{k}_{2}}(\mathbf{r}) =
  \scalebox{0.83}{$
  \left(\  \begin{array}{cc} 
    -\xi \sqrt{3} \kappa &  -\frac{i e^{-\xi i \frac{\pi}{4}}}{\sqrt{2}}  (\partial_{x} - \xi i \partial_{y}) \\
    - \frac{i e^{i \xi \frac{\pi}{4}}}{\sqrt{2}}  (\partial_{x} + \xi i \partial_{y}) & \xi \sqrt{3} \kappa 
    \end{array}\right)
    $}.
     \label{dhr2}
\end{eqnarray}
The block diagonal structure of $\mathcal{\bar{H}}(\mathbf{r})$ reflects the independence of difference Dirac-point sectors.

\subsubsection{Dual vortex and MZM wavefunctions}\label{sec:dualvortex2}
The next step is to introduce an isolated dual vortex in the system in the same way as described in Sect.~\ref{sec:smallJ} by flipping a half-infinite string of links. To find this term, we start from~(\ref{Jstringc}), then use Eq.~(\ref{small_kp_c_l}), introduce a theta function and a delta function as in Eq.~(\ref{Jstringcon}) to find the low energy, long wavelength continuum version of the $J$-string. By doing this, we create a potential $\bar{\mathcal{H}}_\text{$J$-str.}$. Along with the $J$-string just discussed, there is also a $\kappa$-string, which is obtained by flipping the $\kappa$-links along the $x$-axis. In the continuum limit, the total $8\times8$ Hamiltonian $\bar{\mathcal{H}}_{\text{string}} = \bar{\mathcal{H}}_\text{$J$-str.} + \bar{\mathcal{H}}_\text{$\kappa$-str.}$ can also be schematically decomposed as
\[
\bar{\mathcal{H}}_{\text{string}} = \bar{\mathcal{H}}_{\text{string
}}^{\mathrm{nosc}} + \bar{\mathcal{H}}_{\text{string}}^{\mathrm{osc}},
\]
where $\bar{\mathcal{H}}_{\text{string}}^{\mathrm{nosc}}$ contains only the non–oscillatory diagonal entries and 
$\bar{\mathcal{H}}_{\text{string}}^{\mathrm{osc}}$ contains the oscillatory (off-diagonal) terms proportional to 
$e^{2i\mathbf{k}_{1(2)}\cdot \mathbf{r}}$, $e^{i(\mathbf{k}_1-\mathbf{k}_2)\cdot \mathbf{r}}$, etc.
In the end, the string potential is the sum of four terms
\beqn
\mathcal{\bar{H}}_\text{string} 
&=& \mathcal{\bar{H}}^{\text{nosc}}_\text{$J$-str.} + \mathcal{\bar{H}}^{\text{osc}}_\text{$J$-str.} + \mathcal{\bar{H}}^{\text{nosc}}_\text{$\kappa$-str.} + \mathcal{\bar{H}}^{\text{osc}}_\text{$\kappa$-str.},
\eeqn
where the two first are $\propto J$ and the two last $\propto \kappa$. In the following, we discuss the four terms, starting with the two that are non-oscillating.

We first consider the non-oscillating part. 
The diagonal entries of the $8\times 8$ matrix $\bar{\mathcal{H}}_{\text{string}}(\mathbf{r})$ contain only $\pm 1$ and do not depend on the position $\mathbf{r}$. These signs originate from the $J$–string (flipped links) and does not couple the different Dirac points, as they are in different parts of reciprocal space. They encode the branch cut similar to the one discussed in~\ref{subsec:smallJ_Hlink}. In the continuum this produces the ABC
\[
\psi(r,\varphi+2\pi)=-\,\psi(r,\varphi),
\]
which forces the low-energy Dirac equation to admit a normalizable zero–energy 
solution. As will be shown below, these diagonal terms  create two Majorana zero modes,
one in each Dirac-cone pair ($\pm \mathbf{k}_{1(2)}$):
\[
\bar{\mathcal{H}}_{\text{string}}^{\mathrm{nosc}} \, \chi_{\mathbf{k}_1}=0,
\qquad
\bar{\mathcal{H}}_{\text{string}}^{\mathrm{nosc}} \, \chi_{\mathbf{k}_2}=0.
\]
In that direction, our goal is to look for zero energy solutions of $H(\mathbf{r})$~(\ref{kitaevhcl}), in presence of the non-oscillatory part of this $J$-string.
Since the Hamiltonian is block diagonal, we solve for the zero energy solution of~\eqref{dhr1} with ABC
\begin{eqnarray}
    \bar{\mathcal{H}}_{\mathbf{k}_{1}}^{\xi}(\mathbf{r})~\psi_\xi = 0, 
\end{eqnarray}
where $\psi_\xi = \left(\  \begin{array}{cc} 
    \psi_{1}^{\xi}\\
    \psi_{2}^{\xi}
\end{array}\right)$ is a two-component spinor. In polar coordinates, this gives us the pair of equations:
\begin{eqnarray}
  -\xi \sqrt{3} \, \kappa \, \psi_{1}^{\xi} - i \frac{1}{\sqrt{2}} e^{\xi i (\frac{11\pi}{12}-\phi)} (\partial_{r}-\xi i r^{-1} \partial_{\phi}) \psi_{2}^{\xi} = 0
\end{eqnarray}
\begin{eqnarray*}
    \xi \sqrt{3} \, \kappa \, \psi_{2}^{\xi} - i \frac{1}{\sqrt{2}} e^{-\xi i (\frac{11\pi}{12}-\phi)} (\partial_{r}+\xi i r^{-1} \partial_{\phi}) \psi_{1}^{\xi} = 0
\end{eqnarray*}
The following ansatz solves the pair of equations:
\begin{align}
    \psi_{1}^{\xi} & = g_l(r) e^{-\xi i\frac{\phi}{2}}  \\
    \psi_{2}^{\xi} & = -\xi~i~g_l(r) e^{-\xi i \frac{11\pi}{12}} e^{\xi i\frac{\phi}{2}},
    \label{eq:58}
\end{align}
where
\begin{equation}
g_l(r) = \frac{e^{- r/l}}{\sqrt{4 \pi r l}},
\label{eq:g(r)}
\end{equation}
with  the characteristic length $l = 1/(\sqrt{6}\kappa)$. Hence, the complete zero energy solution of~(\ref{dirack1}), constructed on a pair of Dirac cones is
\begin{eqnarray}
    \chi_{\mathbf{k}_{1}} & =  g_l(r)\left(\  \begin{array}{cccc}  
    e^{i(-\frac{\phi}{2}+\mathbf{k}_{1}\cdot\mathbf{r} -\frac{7\pi}{24}+\theta_1)}\\
    e^{i(\frac{\phi}{2} + \mathbf{k}_{1}\cdot\mathbf{r} + \frac{7\pi}{24} + \theta_1 )}\\
    e^{-i(\frac{\phi}{2}+\mathbf{k}_{1}\cdot\mathbf{r} +\frac{7\pi}{24} + \theta_1)}\\
    e^{-i(-\frac{\phi}{2} + \mathbf{k}_{1}\cdot\mathbf{r} - \frac{7\pi}{24} + \theta_1)}
    \end{array}\right),
    \label{evd1}
\end{eqnarray}
where $\theta_1$ is a relative phase between the wavefunctions in the two valleys $+\mathbf{k}_{1}$ and $-\mathbf{k}_{1}$. 

Proceeding in a similar way, we can find the MZM coming from the second pair of Dirac points
\begin{eqnarray}
    \chi_{\mathbf{k}_{2}} & =  g_l(r) \left(\  \begin{array}{cccc}  
    e^{i(-\frac{\phi}{2}+\mathbf{k}_{2}\cdot\mathbf{r} +\frac{\pi}{8}+\theta_2)}\\
    e^{i(\frac{\phi}{2} + \mathbf{k}_{2}\cdot\mathbf{r} - \frac{\pi}{8} + \theta_2 )}\\
    e^{-i(\frac{\phi}{2}+\mathbf{k}_{2}\cdot\mathbf{r} -\frac{\pi}{8} + \theta_2)}\\
    e^{-i(-\frac{\phi}{2} + \mathbf{k}_{2}\cdot\mathbf{r} + \frac{\pi}{8} + \theta_2)}
    \end{array}\right),
    \label{evd2}
\end{eqnarray}
where $\theta_2$ is a phase difference between $+\mathbf{k}_{2}$ and $-\mathbf{k}_{2}$. 

Thus, the two zero-energy solutions of Hamiltonian~(\ref{kitaevhcl}) are:  
\begin{eqnarray}
    \chi_{1} = (\chi_{\mathbf{k}_{1}},0)^{T}  \; \; , \; \; \chi_{2} = (0,\chi_{\mathbf{k}_{2}})^{T}. \label{chif}
\end{eqnarray}
In real space, these two MZMs are bound to the same dual vortex. In reciprocal space, one MZM is located in a pair of Dirac points $\pm \mathbf{k}_1$, while the other is in the other pair $\pm \mathbf{k}_2$.
Unlike (\ref{MZMb}) and (\ref{MZMw}), the wavefunctions $\chi_{\textbf{k}_1}$ and $\chi_{\textbf{k}_2}$ are continuum spinors obtained after projection onto the low-energy subspace near the Dirac points and are expressed in a projected valley basis. Their components no longer directly refer to lattice sites. Consequently, lattice symmetries such as mirror symmetry do not directly constrain the values of $\theta_1$ and $\theta_2$. For this reason, we leave these phases arbitrary within the continuum theory.

The non-oscillating part of the $\kappa$-string $\mathcal{\bar{H}}_\text{$\kappa$-str.}^{\text{nosc}}$ does not couple the two MZMs as it cannot couple different parts of the reciprocal space. Hence, it does not create any splitting. In order to find a splitting, we need to consider the effect of the oscillating terms $\mathcal{\bar{H}}_\text{$J$-str.}^\text{osc}$ and $\mathcal{\bar{H}}_\text{$\kappa$-str.}^\text{osc}$. 

\subsubsection{Energy splitting}\label{sec:energysplitting2}
The solutions~(\ref{chif}) mean that a single vortex-free plaquette in a vortex-full background traps two MZMs coming from two pairs of Dirac points at $\pm \mathbf{k}_{1}$ and $\pm \mathbf{k}_{2}$. Actually, they are bound to the same plaquette and they should therefore be coupled and not remain at zero energy. Hence, we expect that some part of the Hamiltonian will connect these two. 

To find this term, we notice that when creating the MZM, using the ABC on the wavefunction, we have only been considering diagonal elements (the non-oscillatory parts) of this Hamiltonian. Since we expected the MZM to be coming either from Dirac points~$\pm\mathbf{k}_{1} \text{or} \pm\mathbf{k}_{2}$, and did not consider any coupling between $\mathbf{k}_{1}$ or $\mathbf{k}_{2}$. But in the present case, once we substitute~(\ref{small_kp_c_l}) into~(\ref{Jstringc}), we notice that along with the diagonal term (non-oscillatory), we have off-diagonal elements.
These off–diagonal terms $\bar{\mathcal{H}}^{\mathrm{osc}}_{\text{str.}}$ contain all oscillatory contributions:
\[
(\bar{\mathcal{H}}^{\mathrm{osc}}_{\text{str.}})_{ij}\;\propto\;
e^{i\mathbf{k}_{1}\cdot \mathbf{r}},
\; e^{i(\mathbf{k}_1-\mathbf{k}_2)\cdot \mathbf{r}},
\quad i\neq j.
\]
which couple different valleys and therefore couple the two 
Majorana modes of Eq.~(\ref{chif}). This induces an energy splitting
\begin{eqnarray}
2\epsilon = 2\,
\big|\langle\chi_1|\mathcal{\bar{H}}^{\text{osc}}_{\mathrm{str.}}|\chi_2\rangle\big|,
\label{splitt}
\end{eqnarray}
where only the oscillating part of the string potential $\mathcal{\bar{H}}^{\text{osc}}_{\mathrm{str.}} = \mathcal{\bar{H}}^{\text{osc}}_\text{$J$-str.} + \mathcal{\bar{H}}^{\text{osc}}_\text{$\kappa$-str.}$ contributes.
The leading oscillatory contribution to the half-splitting can be written as
\begin{eqnarray}
\epsilon = \big|\langle\chi_1|\mathcal{\bar{H}}^{\mathrm{osc}}_{J\text{-str.}}|\chi_2\rangle\big|
& \simeq &\left|
\int_{\text{string}}\!\!\!\!\!\!\!\!\! d^2\mathbf{r}\; g_l(r)^2 \; f_{\rm osc}(\mathbf{r})\right|\nn \\
& = &\left|\int_b^\infty \!\!\!\!\!\! dx\; g_l(x)^2 \; f_{\rm osc}(x)\right|,
\end{eqnarray}
where \(g_l(r)\) is the MZM radial wavefunction profile \eqref{eq:g(r)} and \(f_{\rm osc}(\mathbf{r})\) encodes the
oscillatory factors (e.g. \(e^{i(\mathbf{k}_1-\mathbf{k}_2)\cdot \mathbf{r}}\)) and the amplitude $J$ of the string potential. In the last step we used the delta function to integrate out $y$. The oscillating function is given by
\begin{equation}
    f_{\rm osc}(x)  = (1+i\sqrt{3})~e^{-2i ( k_{1}^{x} + k_{2}^{x})x}  + 4\sin[2(k_{1}^{x} + k_{2}^{x})x ]
\end{equation}
where $k_{1}^{x}=k_{2}^{x}=\pi/6$. As $f_{\rm osc}(x)$ goes to a constant when $x\to 0^+$, the integrand behaves as
\begin{equation}
g_l(x)^2 = \frac{e^{- 2x/l}}{4 \pi x l}\sim \frac{1}{x},
\end{equation}
and the integral has a logarithmic divergence. The continuum description breaks down at distances of the order of the lattice spacing. Consequently the
matrix element $\big|\langle\chi_1|\mathcal{\bar{H}}^{\mathrm{osc}}_{\mathrm{str.}}|\chi_2\rangle\big|$ requires a short-distance regularization. This cutoff merely encodes unknown vortex core physics (i.e. the detailed form of \(H_{\text{string}}\) inside the unit cell) and does not affect the qualitative scaling of the splitting with \(\kappa\).  We therefore evaluate the radial integral with a lower (UV) cutoff \(b\) chosen to be of the order of the dual vortex core, i.e. half the width of a hexagonal plaquette $a/2=0.5$. As earlier mentioned, the continuum theory does not determine the relative valley phases $\theta_1$ and $\theta_2$. They should be viewed as matching parameters between the continuum and lattice descriptions. In Fig.~\ref{fig:splitkapa}, we show how $\epsilon/\kappa$ depends on the parameters $\theta_1$, $\theta_2$ and $b$. For the reasonable value $b=0.5$, $\theta_1 = \theta_2 = 0$, we compute $\epsilon$ as a function of $\kappa$, see Fig.~\ref{fig:Jversuskappastrings}. At small $\kappa$, $\epsilon$ is linear with a slope $\epsilon\sim 0.6\,\kappa$. This is compatible with the exact result $\epsilon \simeq 0.566~\kappa$ when $\kappa \ll J$ (see below). The precise value of the slope $\epsilon/\kappa$ depends quantitatively on the parameters $\theta_1,\theta_2$, and $b$, but remains of order unity for all reasonable choices considered here.
\begin{figure}[h]
\includegraphics[width=0.9\columnwidth]{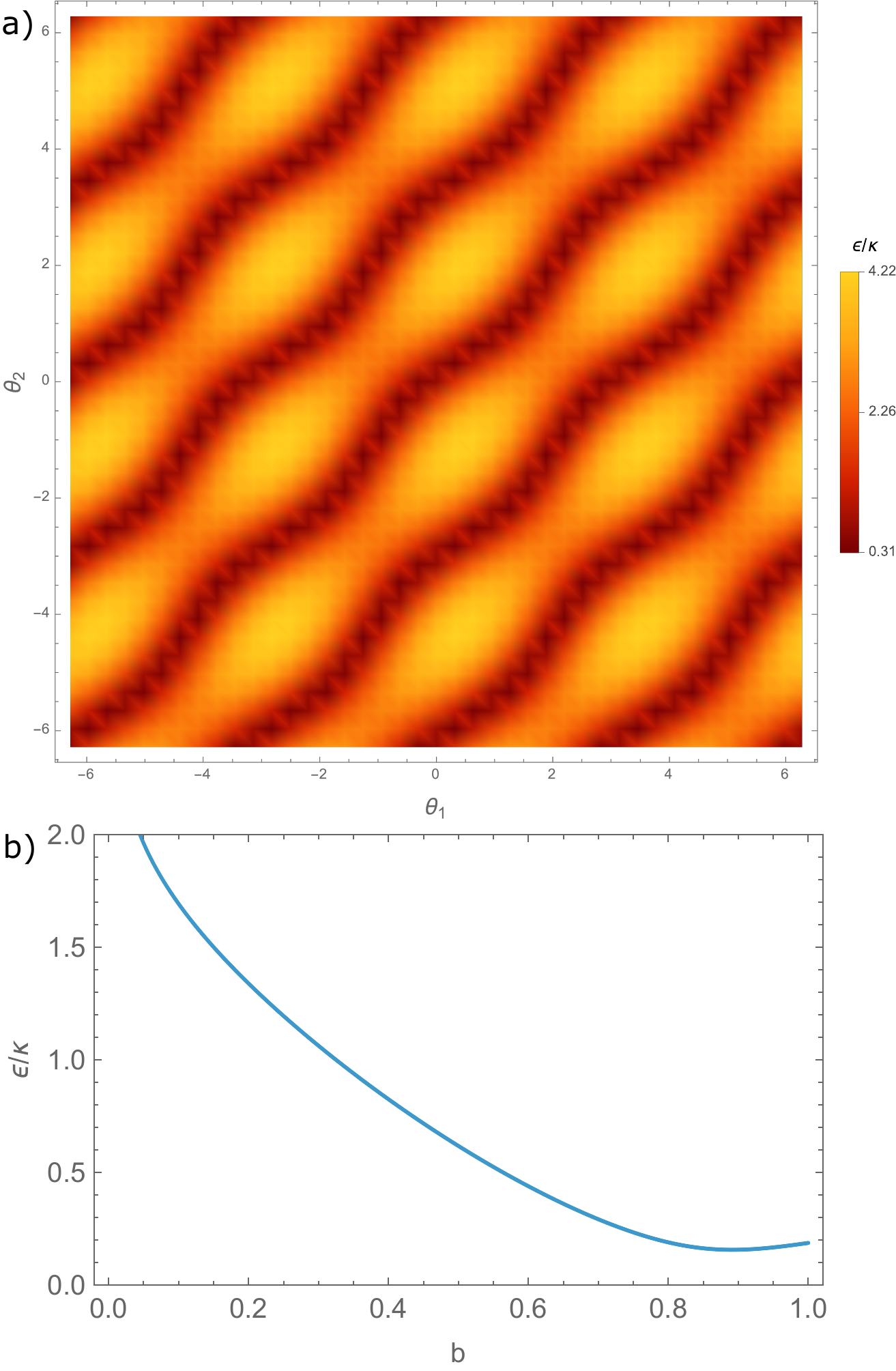}
\caption{\label{fig:splitkapa} 
The slope $\epsilon/\kappa$ at fixed $\kappa=0.01J$ computed using analytical perturbation theory in the continuum and as a function of the UV cutoff $b$ and the phases $\theta_1$ and $\theta_2$: (a) $\epsilon/\kappa$ versus $\theta_1$ and $\theta_2$ at fixed $b=0.5$. (b) $\epsilon/\kappa$ versus $b$ at fixed $\theta_1=\theta_2=0$.}
\end{figure}

A qualitative analytical argument can be made in order to understand why the oscillating $J$-string contributes a splitting of order $\kappa$, while the oscillating $\kappa$-string contributes only at higher order $\kappa^2$. By replacing the oscillating function $f_\text{osc}(x)$ by a constant of order 1, the $J$-string splitting is roughly given by
\beqn \label{eq:JstringOM}
|\langle \chi_1 |\bar{\mathcal{H}}_\text{$J$-str.}^\text{osc}|\chi_2\rangle | &\sim& J\int_\text{str.} \!\!\!\!\!\! d^2r \, g_l(r)^2 = \int_{0.5}^\infty \! dx \, g_l(x)^2  \\
&= & \frac{\sqrt{6}\kappa}{4\pi} \Gamma_0(\sqrt{6}\kappa)
\simeq 0.19 \kappa \ln \frac{1}{\kappa} - 0.29 \kappa, \nn
\eeqn
when $\kappa \to 0$, where $\Gamma_0(x)$ is an incomplete Gamma function. It comes from an operator $H_\text{$J$-str.}^\text{osc}$ of order $\sim J=1$ and a squared MZM wavefunction contribution of order $1/l= \sqrt{6} \kappa$. In this approximation (that neglects oscillating terms in the integrand), the slope $\epsilon/\kappa$ is actually not a constant but has a log behavior at small $\kappa$. This divergence at small $\kappa$ does not exist in the full calculation including $f_\text{osc}(x)$ [see Fig.~\ref{fig:splitkapa}(a)]. For the oscillating $\kappa$-string, the perturbation operator $\bar{\mathcal{H}}_\text{$\kappa$-str.}^\text{osc}$ is already of order $\kappa$. By a similar approximation, the $\kappa$-string splitting can therefore be estimated as
\beqn \label{eq:kappastringOM}
|\langle \chi_1 |\bar{\mathcal{H}}_\text{$\kappa$-str.}^\text{osc}|\chi_2\rangle | &\sim& \kappa \int_\text{str.} \!\!\!\!\!\! d^2r \, g_l(r)^2 
\simeq \frac{\sqrt{6}\kappa^2}{4\pi} \Gamma_0(\sqrt{6}\kappa),
\eeqn
which behaves as $\sim \kappa^2 \ln \frac{1}{\kappa}$ when $\kappa \to 0$.
\begin{figure}[h]
\includegraphics[width=0.9\columnwidth]{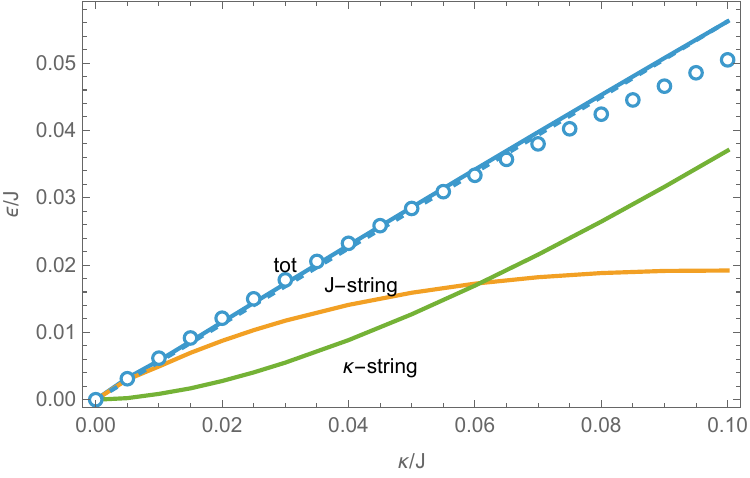}
\caption{\label{fig:Jversuskappastrings}
Full numerics (continuous line, with $36^2$ plaquettes on a torus), numerical perturbation theory (dashed line, barely visible, with $31^2$ plaquettes on a torus) and analytical perturbation theory (open symbols, with $b=0.5$ and $\theta_1=\theta_2=0$) for the half-splitting $\epsilon$ as a function of $\kappa$ (both in units of $J$) in the small $\kappa$ limit. The total result is in blue, the $J$-string contribution in yellow and the $\kappa$-string one in green.}
\end{figure}

\subsection{Numerical perturbation theory}
\label{sec:nptsmallkappa}
We consider the full Hamiltonian with a small finite $\kappa$ (typically $\simeq 0.1 J$) on a torus with an \textit{odd} number ($N_p=31^2=961$) of hexagonal plaquettes and create a single dual vortex (see the corresponding discussion in Sec.~\ref{sec:nptsmallJ}). The smallest value of $\kappa$ that we can reach numerically is related to the system size and the fact that the MZM should be well-localized. As the localization length is $l=1/(\kappa \sqrt{6})$, it is smaller than the system size when $\kappa/J > 1/\sqrt{6N_p}$. 

The two MZM wavefunctions $\chi_1$ and $\chi_2$ are obtained numerically as follows. By diagonalizing the complete Hamiltonian, we find two in-gap states $\psi$ and $\psi^*$ at finite energy $\epsilon >0$ and $-\epsilon <0$. These are the coupled Majorana modes. We first need to decouple them by taking the real or imaginary part of $\psi$. However, there is an unknown global phase $\alpha$ that needs to be taken care of before taking the real or imaginary part. We write $\chi_1= (e^{i\alpha}\psi + e^{-i\alpha}\psi^*)/\sqrt{2}$ and $\chi_2 = (e^{i\alpha}\psi - e^{-i\alpha}\psi^*)(i/\sqrt{2})$ and determine $\alpha$ by asking that $\chi_1(\mathbf{r})$ and $\chi_2(\mathbf{r})$ have inversion symmetry $\chi(-\mathbf{r})=\pm \chi(\mathbf{r})$. These MZM wavefunctions are not eigenvectors of $H$ but have an average energy of zero: $\langle \chi_{1,2} |H|\chi_{1,2}\rangle=0$. They are plotted in Figure~\ref{fig:MZMwfkappa0p1} as blue dots.
\begin{figure}[h]
\includegraphics[width=0.9\columnwidth]{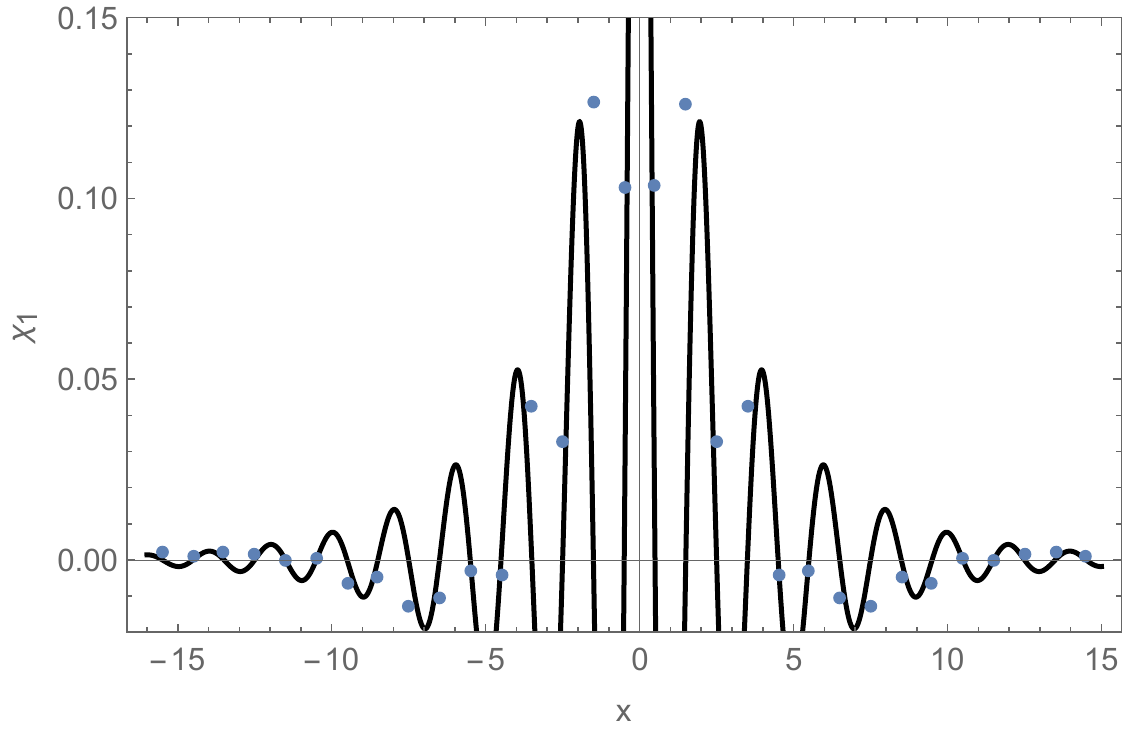}
\includegraphics[width=0.9\columnwidth]{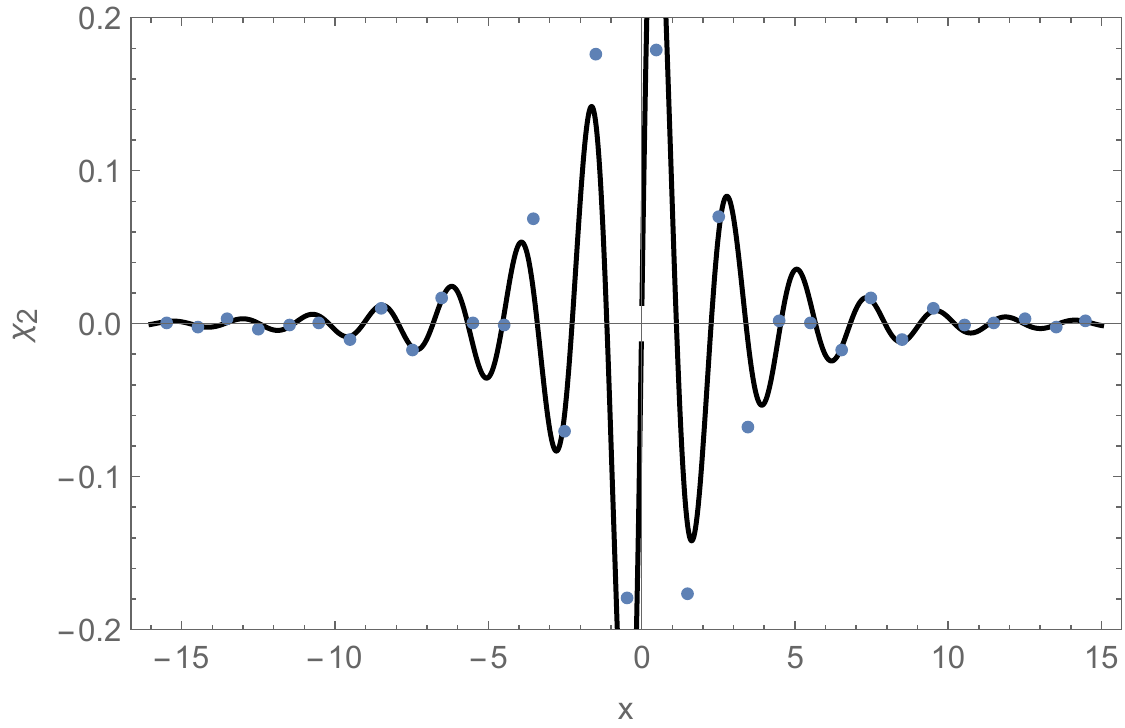}
\caption{\label{fig:MZMwfkappa0p1} Numerically-obtained MZM wavefunctions $\chi_1(x,y)$ and $\chi_2(x,y)$ at fixed $y =1/(2\sqrt{3})$ (just above the dual vortex) for $\kappa=0.1 J$ and $31^2$ plaquettes (blue dots). The black line is $g_l(r)\cos(k_{1,2} \, x +\varphi)$ with $k_1=\pi$ and $\varphi=0$ (up) and $k_2=0.88 \pi$ and $\varphi=-\pi/2$ (down).}
\end{figure}

As in the small $J$ part, we compare the numerical wavefunctions to the following analytical form
\begin{equation} \label{eq:lwfsmallkappa}
\psi(\mathbf{r}) = g_l(r) \cos(\mathbf{k}_{1,2} \cdot \mathbf{r} +\varphi_{1,2})
\end{equation}
that has no branch cut, in contrast to Eqs.~\eqref{MZMb} and \eqref{MZMw}. Inversion symmetry $\psi(-\mathbf{r})=\pm \psi(\mathbf{r})$ imposes that $\varphi_{1,2}=0$ mod $\pi/2$. In Fig.~\ref{fig:MZMwfkappa0p1}, we compare the numerically-obtained wavefunctions with Eq.~\eqref{eq:lwfsmallkappa} in which $\mathbf{k}_{1,2}$ and $\varphi_{1,2}$ are treated as fitting parameters. Because the numerical and analytical gauges are different and also because we used a continuum limit in the analytical calculation, we can not expect a perfect match between the two sets of wavefunctions. 
 

We numerically find that $\chi_1$ and $\chi_2$ are coupled by 
\begin{equation}
\langle \chi_2| H |\chi_1\rangle \simeq i\, 0.562\, \kappa ,
\label{155}
\end{equation}
so that the half-splitting 
$\epsilon\simeq  0.562 \, \kappa$. 
The comparison with analytical perturbation theory in the continuum limit and the full numerics (see next section) is shown in Fig.~\ref{fig:Jversuskappastrings}.

\subsection{Full numerics}\label{sec:fnsmallkappa}
We have also performed a fully numerical (non-perturbative) calculation of the splitting as a function of $\kappa$. Here we choose to work with an \textit{even} number of plaquettes ($36^2=1296$) and create a pair of dual vortices that are then placed far apart (about 18 plaquettes away, i.e. half of the linear system size). 

One first need to identify the string potential due to flipping the links needed to create the dual vortex. Then it is possible to separate this potential into a contribution proportional to $J$ (this is the $J$-string potential) and a contribution proportional to $\kappa$ (i.e. the $\kappa$-string potential). At low $\kappa$, the first produces a linear in $\kappa$ splitting (see the yellow line in Fig.~\ref{fig:Jversuskappastrings}), while the second produces a quadratic in $\kappa$ splitting (see the green line in Fig.~\ref{fig:Jversuskappastrings}). This is in qualitative agreement with analytical perturbation theory, see \eqref{eq:JstringOM} and \eqref{eq:kappastringOM}. At larger $\kappa$, the $J$-string contribution saturates. A surprising feature is that the total contribution remains linear on a large range of $\kappa$ despite the fact that the two individual contributions are not linear. Indeed, when $0.1<\kappa/J< 0.3$, $\epsilon_\text{$J$-string}\simeq 0.02 J$ and $\epsilon_\text{$\kappa$-string}\simeq 0.56\kappa-0.02 J$. 

We find that the total half-splitting $\epsilon\simeq 0.566 \kappa$ in the $\kappa \ll J$ limit, very close to the result found by numerical perturbation theory result (see previous section).

 \subsection{Conclusion}
We recap the contributions of different terms in the Hamiltonian:
 
(i) The bulk $J$ term is responsible for producing a dispersion relation with 4 Dirac points (that separate in two pairs), i.e. 4 gapless Dirac cones. The bulk $\kappa$ term opens a small gap $\Delta \simeq 2\sqrt{3}\kappa$ in the 4 Dirac cones. The system is then equivalent to a topological (chiral) superconductor with Chern number $+ 2$~\cite{Lahtinen10}.
 
Then introducing a dual vortex via a semi-infinite string creates a potential, that contains an oscillating and a non-oscillating part:
 
(ii) The non-oscillating part of the $J$-string creates the branch cut responsible for the appearance of two MZMs per dual vortex. Each MZM resides in a pair of Dirac points. The non-oscillating part of the $\kappa$-string leaves the MZMs unaffected.

(iii) The energy splitting is due to the coupling between MZMs within a dual vortex and is provided by oscillating terms of the string potential. The oscillating $J$-string produces a splitting $\propto \kappa$, while the oscillating $\kappa$-string produces a splitting $\propto \kappa^2$, see Fig.~\ref{fig:Jversuskappastrings}.


\section{Analogy to a chiral superconductor}\label{sec:anachiralsupercond}
In the ten-fold classification of topological bands, the Kitaev honeycomb model belongs to the D class~\cite{Kitaev09}. The latter also hosts chiral superconductors. In the present section we compare the Kitaev model in the vortex-full sector with a chiral superconductor with $\pm 2$ Chern number and concentrate on the states bound to a ``vortex'' in both cases.

Caroli-de Gennes-Matricon states are mid-gap states bound to the core of a 3D superconducting vortex~\cite{Caroli64}. In a 2D chiral superconductor characterized by a Chern number $\nu$, these bound states also exist but it is known that there is an interesting even-odd effect~\cite{Volovik99}. When $\nu$ is even (for example in an $s$-wave superconductor with $\nu=0$ or in a $d_{x^2-y^2}\pm id_{xy}$-wave superconductor with $\nu=\pm 2$, see e.g.~\cite{Lee16}), one expects
\begin{equation}
E_n = \epsilon_0 \, (n+1/2)
\end{equation}
with integer $n$ so that the lowest energy states ($n=0$ and $-1$) are at finite energy $\pm \epsilon_0/2$ and there is no zero-energy state. When $\nu$  is odd (for example in a $p_x\pm ip_y$-wave or in a $d_{xz}\pm i d_{yz}$-wave superconductor with $\nu=\pm 1$), one rather expects 
\begin{equation}
E_n = \epsilon_0 \, n,
\end{equation}
featuring, in particular, a zero-energy MZM when $n=0$. 

Independent of the parity of the Chern number, the energy splitting between successive bound states is $\epsilon_0$ and is estimated as $\epsilon_0\sim \Delta^2/E_F$ in weak coupling~\cite{Volovik99} (i.e. when $\Delta/E_F \ll 1$) and as $\epsilon_0\sim\Delta$ in strong coupling~\cite{Sau10} (i.e. when $\Delta/E_F \gg 1$), where $E_F$ is the Fermi energy in the absence of superconductivity and $\Delta$ is the superconducting pairing gap. In weak coupling, one therefore expects many equidistant mid-gap states bound to the vortex core, because the splitting is small compared to the bulk gap. Whereas in strong coupling, one only expects two mid-gap states, as the splitting is comparable to the bulk gap.

The vortex-full Kitaev honeycomb model (gapped by a three-spin term $\kappa$) has an even Chern number (either $+2$ or $-2$ depending on the ratio $\kappa/J$). We therefore expect that a dual vortex traps mid-gap states at finite energy. This is indeed what we find. Also, it is well-known that in the vortex-free sector, the Chern number is $\pm 1$ and that an isolated vortex traps a single MZM~\cite{Kitaev06,Otten19}. These two facts agree with the even/odd effect found by Volovik.

However, the vortex-full Kitaev honeycomb model also corresponds to rather strong-coupling superconductivity when $J\ll \kappa$ (as $\Delta \sim \kappa$ and $E_F\sim \kappa$ estimated as the total bandwidth) and to weak-coupling superconductivity when $\kappa \ll J$ (as $\Delta \sim \kappa$ and $E_F\sim J$ estimated as the total bandwidth). For the mid-gap states bound to a dual vortex, we would therefore expect only one or two (depending on the Chern number parity) states with a splitting $\epsilon_0\sim \kappa$ at small $J$ and many states with a splitting $\epsilon_0\sim \kappa^2/J$ at small $\kappa$.

But this is not what we find. (i) First, in both limits, we only see a pair of states at energy $\pm \epsilon$ in the bulk gap (and not a whole ladder of states). (ii) Second, we find $\epsilon \sim J$ (and not $\epsilon_0 \sim \kappa$) at small $J$ and $\epsilon \sim \kappa$ (and not $\epsilon_0 \sim \kappa^2/J$) at small $\kappa$.

This shows that, despite the analogy and the agreement with the even/odd effect, the behavior of the bound states to a dual vortex in the Kitaev honeycomb model is quite different from that in a chiral superconductor.

\section{Conclusion and perspectives}
\label{sec:conclusion}
In this paper, we have computed the energy splitting of the pair of Majorana modes trapped in a dual vortex of the vortex-full sector of the Kitaev honeycomb model, in the two limits of small $\kappa$ and small $J$. Our findings are summarized in Table~\ref{table:1}.
\begin{table}[h!]
\centering

\begin{tabular}{ c || c| c  } 
    &    small $J$  &  small $\kappa$   \\
\hline
\hline
Gapped background:    &  $H_{\kappa-\text{bulk}}$   &  $H_{\text{bulk}}$   \\
Chern number    &  $\nu=-2$   &  $\nu=+2$   \\
Bulk gap             & $\Delta \simeq 2\sqrt{3}\kappa$ & $\Delta \simeq 2\sqrt{3}\kappa$ \\
\hline
Branch cut:   &  $H_{\kappa-\text{string}}$   &   $H_{\text{string}}^\text{nosc}$  \\
MZM wavefunction & $g(r)\cos(k_+ x +\varphi)$ &$g_l(r)\cos(k_{1,2} x +\varphi)$  \\
\hline
Perturbation: & & \\
 zero mode coupling    &  $H_{J}$   &   $H_{\text{string}}^\text{osc}$  \\
\hline
Half-splitting: & & \\
Analytical pertur. th.    &  $\epsilon \simeq 0.57 J$   &   $\epsilon \sim 0.6 \kappa$ (*) \\
Numerical pertur. th.   &   $\epsilon \simeq 0.3933 J$  &   $\epsilon \simeq 0.562 \kappa$  
\\
Full numerics   &  $\epsilon \simeq 0.3933 J$   &  
$\epsilon \simeq 0.566 \kappa$   \\
\end{tabular}

\caption{Different Hamiltonian terms responsible for (i) making the gapped vortex-full background, (ii) creating a dual vortex binding a pair of MZMs and (iii) coupling them, that leads to a splitting. Chern numbers $\nu$, bulk gaps $\Delta$, MZM wavefunctions and half-splittings $\epsilon$ are also indicated. (*) For $b = 0.5$ and $\theta_1 = \theta_2 = 0$.}
\label{table:1}
\end{table}

When $J = 0$, the problem splits up into two copies of the Claro-Wannier model of a triangular lattice in a uniform magnetic field~\cite{Claro79}. In this case, we found two MZMs at the same spatial location of the dual vortex. As the MZMs belong to two different triangular sublattices, they do not interact. Once we switch on a small $J$, it couples the two MZMs and lifts the degeneracy, leading to a half-splitting $\epsilon \sim J$. 

Similarly, for $\kappa\ll J$, we identified two parts of the Hamiltonian, one (the non-oscillating $J$-string) which creates the MZMs, and the other part (the oscillating $J$-string) which couples these two. In this limit, the separation of the Hamiltonian into a part that produces MZMs and a part that couples them is less obvious. Also, the splitting is found to be $\epsilon \sim \kappa$ but is obtained from a perturbation $\propto J$.

As perspectives, it would be interesting to study the vicinity of $\kappa=J/2$, corresponding to a bulk gap closing of the vortex-full sector. It corresponds to a topological phase transition at which the Chern number jumps from $+2$ to $-2$. One could also compute the interaction between two such dual vortices, similarly to what was done in the vortex-free case~\cite{Otten19}. At a large distance, the main effect is the on-site coupling between the two Majorana modes trapped on the same vortex. But at shorter distances, the two vortices start interacting and there is also a tunneling probability between the vortices. This should be compared with the corresponding calculation done in the case of two vortices in the vortex-free sector~\cite{Lahtinen12, Cheng09}. This tunneling is an essential ingredient of the effective models describing the dual vortex lattices studied in~\cite{Alspaugh24} and that gives rise to almost all of the 16 phases predicted by Kitaev~\cite{Kitaev06}. Another interesting direction would be to study the influence of real space curvature on the states bound to a vortex (or a dual vortex) by considering hyperbolic lattices.

\acknowledgements
 We thank David Alspaugh, Andrzej Ptok, Przemysław Piekarz for discussions and Julien Vidal for many useful discussions and for reading a preliminary version of this article. S. B. acknowledges LPTMC Paris, where this work was started, for hospitality.
 S.B. also acknowledges the financial support provided by the Polish National Agency for Academic Exchange NAWA  under the Programme STER– Internationalisation of Doctoral Schools, Project no. PPI/STE/2020/1/00020.

\appendix


\section{Vortex-full model}
\label{app:vortexfulldr}
The vortex-full sector for the Kitaev model on the honeycomb lattice was studied in~\cite{Lahtinen08,Lahtinen10}, see also Appendix C in~\cite{Fuchs20}. The Hamiltonian, written in canonical basis and using the unit cell shown in Fig.~\ref{fig.carlo_vortex}(a), reads
\begin{eqnarray}
    H(\mathbf{k}) = \sum_{\mathbf{k}} \mathbf{\Psi}^{\dagger}_{\mathbf{k}}~\mathcal{H}(\mathbf{k})~\mathbf{\Psi}_{\mathbf{k}}.\label{kdual}
\end{eqnarray}
where $\mathbf{\Psi}^{T}_{\mathbf{k}} = (\chi_{\mathbf{k}}^A,\chi_{\mathbf{k}}^B,\chi_{\mathbf{k}}^C,\chi_{\mathbf{k}}^D)^{T}$ and
\begin{eqnarray}
\begin{aligned}
\mathcal{H}(\mathbf{k}) 
= {} & iJ 
\begin{pmatrix}
0 & \mathcal{H}_{\boldsymbol{\delta}} \\
-\mathcal{H}_{\boldsymbol{\delta}}^{*} & 0
\end{pmatrix} \\
& {} + 
i\kappa
\begin{pmatrix}
\mathcal{H}_{\mathbf{n}_{23}} + \mathcal{H}_{\mathbf{n}_{1}} & 0 \\
0 & -\mathcal{H}_{\mathbf{n}_{23}} + \mathcal{H}_{\mathbf{n}_{1}}
\end{pmatrix}
\end{aligned}\label{A:kitaevh}
\end{eqnarray}
$\mathcal{H}_{\boldsymbol{\delta}}, \mathcal{H}_{\mathbf{n}_{23}}, \mathcal{H}_{\mathbf{n}_{1}}$ are all functions of $\mathbf{k}$, explicitly given by
\begin{eqnarray*}
       \mathcal{H}_{\boldsymbol{\delta}} = \left(\  \begin{array}{cc} -e^{i\mathbf{k}\cdot\boldsymbol{\delta}_{1}} - e^{i\mathbf{k}\cdot\boldsymbol{\delta}_{3}} & -e^{i\mathbf{k}\cdot\boldsymbol{\delta}_{2}}\\
    -e^{i\mathbf{k}\cdot\boldsymbol{\delta}_{2}} & e^{i\mathbf{k}\cdot\boldsymbol{\delta}_{1}} - e^{i\mathbf{k}\cdot\boldsymbol{\delta}_{3}} \end{array}\right),
\end{eqnarray*}
\begin{eqnarray*}
       \mathcal{H}_{\mathbf{n}_{23}} =  \left(\  \begin{array}{cc} e^{-i\mathbf{k}\cdot\mathbf{n}_{2}} - e^{i\mathbf{k}\cdot\mathbf{n}_{2}} & - e^{i\mathbf{k}\cdot\mathbf{n}_{3}} + e^{-i\mathbf{k}\cdot\mathbf{n}_{3}}\\
    e^{-i\mathbf{k}\cdot\mathbf{n}_{3}} - e^{i\mathbf{k}\cdot\mathbf{n}_{3}} & - e^{-i\mathbf{k}\cdot\mathbf{n}_{2}} + e^{i\mathbf{k}\cdot\mathbf{n}_{2}} \end{array}\right),
\end{eqnarray*}
and
\begin{eqnarray*}
       \mathcal{H}_{\mathbf{n}_{1}} =  \left(\  \begin{array}{cc} 0 & e^{i\mathbf{k}\cdot\mathbf{n}_{1}} + e^{-i\mathbf{k}\cdot\mathbf{n}_{1}}\\
   -e^{i\mathbf{k}\cdot\mathbf{n}_{1}} - e^{-i\mathbf{k}\cdot\mathbf{n}_{1}} & 0 \end{array}\right)
\end{eqnarray*}
with the vectors $\boldsymbol{\delta}_{1} = (0,\frac{1}{\sqrt{3}}), \boldsymbol{\delta}_{2} = (-\frac{1}{2},-\frac{1}{2\sqrt{3}}), \boldsymbol{\delta}_{3} = (\frac{1}{2},-\frac{1}{2\sqrt{3}})$, while $\mathbf{n}_{1},\mathbf{n}_{2},\mathbf{n}_{3}$ were defined in earlier in the main text.

We can diagonalize the above Hamiltonian to obtain the four bands of the vortex full model as $\pm E_{+}(\mathbf{k}), \pm E_{-}(\mathbf{k})$, where
\begin{eqnarray}\label{eq:dispvf}
    E_{\pm}  = \sqrt{f(\mathbf{k}) \pm 2\sqrt{g(\mathbf{k})}}
\end{eqnarray}
with
\onecolumngrid

\medskip
\begin{eqnarray}
    f(\mathbf{k}) 
    &=& 3 J^2 + 4 \kappa^2[\sin^2(\mathbf{k}\cdot\mathbf{n}_{3}) + \sin^2(\mathbf{k}\cdot\mathbf{n}_{2}) + \cos^2(\mathbf{k}\cdot\mathbf{n}_{1}) ]
\end{eqnarray}
and
\begin{align}
    g(\mathbf{k}) & =  J^4 [\cos^2{(\mathbf{k}\cdot \mathbf{n}_{3})} + \cos^2{(\mathbf{k}\cdot \mathbf{n}_{2})} + \sin^2{(\mathbf{k}\cdot \mathbf{n}_{1})}] + 4 \kappa^2 J^2 \{4[\sin^2{(\mathbf{k}\cdot \mathbf{n}_{3})} + \sin^2{(\mathbf{k}\cdot \mathbf{n}_{2})} + \cos^2{(\mathbf{k}\cdot \mathbf{n}_{1})}]-3\}
\end{align}
\twocolumngrid
Compared to equations (C3), (C4) and (C5) in the Appendix C of~\cite{Fuchs20}, the above equations only differ by a gauge choice.

\section{Small $J$ limit}
\label{app:smallJ}
\subsection{Derivation of MZM wavefunction}
In this appendix, we derive the wavefunction~(\ref{carlomzm}), which is the solution of~(\ref{mzm_smallj}), with ABC. Eq.~\eqref{mzm_smallj} gives us the pair of equations
\begin{eqnarray}
\xi~\psi^{A}_{\xi} -i~ \{(1-\xi\frac{i}{2}) \partial_{x} - i ~\xi\frac{\sqrt{3}}{2}\partial_{y}\}~\psi^{B}_{\xi} &=& 0 \label{carloMZM1}\nn \\
i~ \{(1+\xi\frac{i}{2}) \partial_{x} + i ~\xi\frac{\sqrt{3}}{2}\partial_{y}\} \psi^{A}_{\xi}
+ \xi~\psi^{B}_{\xi} &=& 0,
\label{carloMZM2}
\end{eqnarray}
for $\psi^{A}_{\xi}(x,y)$ and $\psi^{B}_{\xi}(x,y)$. To solve them, we change the variables as follows
\begin{eqnarray}
x' &=& x - \frac{1}{\sqrt{3}}y\nn \\
y' &=& \frac{2}{\sqrt{3}}y, 
\label{primed}
\end{eqnarray}
which implies that $\partial_{x}\to \partial_{x'}$ and $\partial_{x}+\sqrt{3}\partial_y \to 2\partial_{y'}$. This leads to 
\begin{eqnarray}
    \xi~\psi^{A}_{\xi} - i~(\partial_{x'}- i \xi~ \partial_{y'})\psi^{B}_{\xi}  & =& 0\nn  \\
    i~(\partial_{x'}+i \xi~ \partial_{y'})\psi^{A}_{\xi} +\xi\psi^{B}_{\xi} & =& 0,
\end{eqnarray}
for $\psi^{A}_{\xi}(x',y')$ and $\psi^{B}_{\xi}(x',y')$. In polar coordinates $r' = \sqrt{x'^2 + y'^2}$, $\phi' = \tan^{-1}{(\frac{y'}{x'})}$, they can be rewritten as
\begin{eqnarray}
    \xi~\psi^{A}_{\xi} - i~e^{-i\xi \phi'}(\partial_{r'}- i \frac{\xi}{r'}~ \partial_{\phi'})\psi^{B}_{\xi}  & =& 0\nn  \\
    i~e^{i\xi \phi'}(\partial_{r'}+ i \frac{\xi}{r'}~ \partial_{\phi'})\psi^{A}_{\xi} +\xi\psi^{B}_{\xi} & =& 0,
\end{eqnarray}
for $\psi^{A}_{\xi}(r',\phi')$ and $\psi^{B}_{\xi}(r',\phi')$. Solving these gives us the wavefunction written in~(\ref{carlomzm}).

\subsection{Exact form of the $J$-bulk Hamiltonian}
\label{app:smalljham}
Here we show the exact form of Hamiltonian $H_{J\text{-bulk}}(\mathbf r)$~(\ref{eq:HJ-cont}). The $8\times 8$ continuum Hamiltonian $H_{J\text{-bulk}}(\mathbf r)$ naturally decomposes into 
four $4\times 4$ blocks,
\begin{equation}
\mathcal{H}_{J\text{-bulk}}(\mathbf r)
=
iJ \begin{pmatrix}
0 & H_J^{\mathrm{UR}}(\mathbf r) \\
H_J^{\mathrm{LL}}(\mathbf r) & 0
\end{pmatrix},
\label{eq:HJ-block}
\end{equation}
because the $J$ dependent term always couples the black triangular lattice 
$(A,B)$ to the white triangular lattice $(C,D)$.  
Consequently:
\begin{itemize}
    \item the \emph{upper-left} block acts entirely within the black triangular lattice and therefore cannot couple the two Majorana zero modes;
    \item the \emph{lower-right} block acts within the white triangular lattice and likewise does not contribute to the splitting;
    \item the \emph{upper-right and lower--left block} $H_J^{\mathrm{UR}}(\mathbf r)$ and $H_J^{\mathrm{LL}}(\mathbf r)$ couple the black and white sublattices and are
    responsible for generating the energy splitting.
\end{itemize}
Due to the special structure of the MZM wavefunctions~(\ref{mzmwb}), we find that the half-splitting is controlled entirely by the matrix element of the upper--right block,
\begin{equation}
\epsilon_{J\text{-bulk}} = |\langle \chi_1| \mathcal{H}_\text{$J$-bulk} |\chi_2\rangle| = J|\langle \psi_b| H_\text{$J$}^{UR} |\psi_w\rangle|
\end{equation}
$H^{UR}_{J}$ can be found out to be
\begin{equation}
H^{UR}_{J}
=
\begin{pmatrix}
H_{J_{1}} & 0 \\
0 & H_{J_{2}}
\end{pmatrix},
\end{equation}
where 
\begin{equation*}
H_{J_{1}}
=
\begin{pmatrix}
-p~h_{J_{1}} & p~h_{J_{2}}\\
p~h_{J_{2}}& p~h^{*}_{J_{1}}
\end{pmatrix}
\end{equation*}
and 
\begin{equation}
H_{J_{2}}
=
\begin{pmatrix}
 p~h^{*}_{J_{1}} & p~h_{J_{2}}\\
p~h_{J_{2}}& -p~h_{J_{1}}
\end{pmatrix},
\end{equation}
with $p = \frac{1}{6}(\frac{i}{2}-\frac{\sqrt{3}}{2}),~h_{J1} = (6 - 6i) -3~i~\partial_{x} + (1-2i)\sqrt{3}~i~\partial_{y}$, $h_{J2} = 6i -3~i~\partial_{x} - \sqrt{3}~i~\partial_{y}$. 

\section{Effective Hamiltonian for small $\kappa$}
\label{app:smallkappa}
In this Appendix, we derive the low energy effective Hamiltonian in the small $\kappa$ limit~(\ref{kitaevhcl}), starting from the lattice model. We proceed in the following steps:
\begin{enumerate}
    \item We identify the low energy subspace associated with each Dirac points at $\kappa=0$, and project the full Hamiltonian onto these low energy bands.
    \item We expand the projected Hamiltonian to linear order in momentum at each Dirac point and obtain Dirac like Hamiltonians in momentum space.
    \item We then obtain the final real space Hamiltonian by the substitution $\mathbf{q}\to -i \boldsymbol{\partial}$.
    \item In the end, we show an alternative route by introducing the slowly varying fermion fields and explicitly deriving relations between the lattice operators and the low energy fields.
\end{enumerate}

\subsection{Direct derivation}

We start from the Bloch Hamiltonian of the vortex-full model $H(\mathbf{k})$, given in Appendix~\ref{app:vortexfulldr}. At $\kappa=0$, the spectrum exhibits four Dirac points $\textbf{Q} \in \{\pm\mathbf{k}_{1}, \pm\mathbf{k}_{2}\}$, where two bands touch the zero energy. At each Dirac point $\textbf{Q}$, we diagonalize the Hamiltonian $H(\mathbf{k})$ at $\kappa = 0$
\begin{eqnarray}
    U^{\dagger}_{\textbf{Q}} H(\textbf{Q})U_{\textbf{Q}} = \text{diag}(E_{h1}, E_{h2},0,0)
\end{eqnarray}
where $U_{\textbf{Q}}$ is an unitary matrix whose colums define the eigenvectors of high-energy ($h$) and low-energy ($l$) bands.
In this basis, the Hamiltonian (for arbitrary $\kappa$) near $\mathbf{Q}$ takes the Bloch form
\begin{eqnarray}
\mathcal{H}(\mathbf{k}) = 
\begin{pmatrix} 
    \mathcal{H}_{hh}(\mathbf{k}) &  
    \mathcal{H}_{hl}(\mathbf{k}) \\
    \mathcal{H}_{lh}(\mathbf{k}) & 
    \mathcal{H}_{ll}(\mathbf{k}) 
\end{pmatrix}
\end{eqnarray}
Here $\mathcal{H}_{xx}(\mathbf{k})$ are themselves $2\times2$ matrices. We project onto the low energy subspace by neglecting $ \mathcal{H}_{hh}(\mathbf{k}), \mathcal{H}_{hl}(\mathbf{k}), \mathcal{H}_{lh}(\mathbf{k})$ to leading order. This yields the effective Hamiltonian as 
\begin{eqnarray}
    \mathcal{H}_{eff}(\textbf{k}) \simeq \mathcal{H}_{ll}(\textbf{k})
\end{eqnarray}
This result is valid up to first  order in perturbation theory. Corrections occur at second order in perturbation theory as discussed, e.g., in Appendix B of~\cite{Lim20}.

We then expand the projected Hamiltonian around each Dirac point, $\mathbf{k} = \mathbf{Q} + \mathbf{q}$. Keeping terms to linear order in $\textbf{q}$, we obtain effective Hamiltonians  at the four Dirac points:
\begin{eqnarray}
\mathcal{H}_{\xi \mathbf{k}_{1}}(\mathbf{q}) = 
\scalebox{0.85}{$
\begin{pmatrix} 
    -\xi \sqrt{3} \kappa &  
    \tfrac{1}{\sqrt{2}} e^{\xi i \tfrac{11\pi}{12}} (q_{x} - \xi i q_{y}) \\
    \tfrac{1}{\sqrt{2}} e^{-\xi i \tfrac{11\pi}{12}} (q_{x} + \xi i q_{y}) & 
    \xi\sqrt{3} \kappa  
\end{pmatrix}
$} \label{hq1}
\end{eqnarray}
and
\begin{eqnarray}
\mathcal{H}_{\xi \mathbf{k}_{2}}(\mathbf{q}) = 
\scalebox{0.85}{$
\begin{pmatrix}
   -\xi \sqrt{3} \kappa &
   \tfrac{1}{\sqrt{2}} e^{-\xi i \tfrac{\pi}{4}} (q_{x} - \xi i q_{y}) \\
   \tfrac{1}{\sqrt{2}} e^{\xi i \tfrac{\pi}{4}} (q_{x} + \xi i q_{y}) &
   \xi \sqrt{3} \kappa
\end{pmatrix}
$} \label{hq2}
\end{eqnarray}
where $\xi=\pm1$ is a valley index and we use energy units such that $J=1$ in this section.
Finally, the continuum limit is obtained by the substituion $\mathbf{q}\to -i \boldsymbol{\partial}$ in the Bloch Hamiltonians~(\ref{hq1}, \ref{hq2}). This yields the final low-energy and long-wavelength real space Dirac Hamiltonians~(\ref{dhr1}, \ref{dhr2}), which constitute the building blocks for the full Hamiltonian~(\ref{kitaevhcl}). The Hamiltonians~(\ref{hq1}, \ref{hq2}) describe the low-energy degrees of freedom in the vicinity of Dirac points and are written in a two component basis corresponding to low energy bands. For each Dirac point $\textbf{k}_{i}$ ($i \in \{1,2\}$), there are two valleys labeled by $\xi = \pm$, corresponding to expansions around $\pm \textbf{k}_{i}$. These can be combined into a four-component structure as written in~\eqref{dirack1}. Finally, since the low-energy sectors associated with $\textbf{k}_{1}$ and $\textbf{k}_{2}$ are well separated in momentum space, they remain independent and can be combined into a enlarged spinor, leading to a block diagonal spinor written in~(\ref{hfull}). Collecting all contributions yields the final Hamiltonain written in~(\ref{kitaevhcl}).

\subsection{Lattice fermion operator to continuum field mapping}
The above method does not specify how the original lattice fermion operators are related to the low energy degrees of freedom. Hence, we now construct the continuum theory more systematically by introducing slowly varying fermion fields around each Dirac point. This route is equivalent to the direct substitution above, but provides a operator-level mapping which is used in deriving the string terms.

\subsubsection{Dirac-point dependent basis}
A crucial point is that the low-energy subspace is defined separately at each Dirac point. At each Dirac point $\textbf{Q}$, the Bloch Hamiltonian is diagonalized using a unitary matrix $U_{\textbf{Q}}$, defining a new basis:
\begin{eqnarray}
    \ket{\text{new}}_{\mathbf{Q}} = U^{\dagger}_{\mathbf{Q}} \, \ket{\text{old}} \label{basis_trans}
\end{eqnarray}
In this new basis, the original fermion operator splits into i) two low energy modes $l^{\mathbf{Q}}_{1,2}$ , and ii) two high energy modes $h^{\mathbf{Q}}_{1,2}$. Importantly, the transformation $U_{\textbf{Q}}$ depends on $\textbf{Q}$. Hence, the decomposition of the original lattice operators $\mathbf{\Psi}^{T}_{\mathbf{k}} = (\chi_{\mathbf{k}}^A,\chi_{\mathbf{k}}^B,\chi_{\mathbf{k}}^C,\chi_{\mathbf{k}}^D)^{T}$, in terms of the low and high energy fields~(\ref{basis_trans}) are different for different $\textbf{Q}$.
For example, for sublattice A, near $+\textbf{k}_{1}$:
\begin{eqnarray}
\chi^A(\mathbf{+k}_{1}) =  \lambda \, l_{2}^{+\mathbf{k}_{1}} + \mu \, l_{1}^{+\mathbf{k}_{1}} +  \zeta \, h_{2}^{+\mathbf{k}_{1}} + \eta \, h_{1}^{+\mathbf{k}_{1}}
\end{eqnarray}
where $\lambda$, $\mu$, $\zeta$, $\eta$ are some complex coefficients, determined by $U_{+\textbf{k}_{1}}$. Projecting onto the low-energy subspace amounts to negtecting the high-energy operators $h^{+\textbf{k}_{1}}_{1,2}$:
\begin{eqnarray}
\chi_{\mathbf{+k}_{1}}^A \simeq  \lambda \, l_{2}^{+\mathbf{k}_{1}} + \mu \, l_{1}^{+\mathbf{k}_{1}}
\end{eqnarray}
Analogous relations hold at other Dirac points $-\textbf{k}_{1}, \pm \textbf{k}_{2}$, but with different complex coefficients.

\subsubsection{Slowly varying continuum fields} 
We can now define the operators $l_{i}(\mathbf{k}_{1}+\mathbf{q})$ which annihilates a fermion in the $i$-th low-energy band near $\mathbf{k}_1$, as the smooth continuation of the low-energy eigenmodes  $l_{i}^{+\mathbf{k}_{1}}$ of  $H(\mathbf{k}_{1})$.
\begin{eqnarray}
\chi^A(\mathbf{k}_{1} + \mathbf{q}) \simeq  \lambda \, l_{2}(\mathbf{k}_{1} + \mathbf{q}) + \mu \, l_{1}(\mathbf{k}_{1} + \mathbf{q})
\label{eq:projected_chi}
\end{eqnarray}
Low-energy physics is governed by momenta in the vicinity of the Dirac point
$\mathbf{k} = \mathbf{k}_1 + \mathbf{q}$, with $|\mathbf{q}| \ll |\mathbf{k}_1|$. We define slowly varying continuum fermion
fields by Fourier transforming only the residual momenta $\mathbf{q}$
\begin{eqnarray}
    L_{i}^{+\mathbf{k}_{1}}(\mathbf{r}) \equiv \int_{|q|\ll 1}
\frac{d^2 q}{(2\pi)^2} \, e^{i \mathbf{q}\cdot \mathbf{r}} \, l_i(\mathbf{k_1 + q}).
\label{eq:continuum_fields}
\end{eqnarray}

\subsubsection{Reconstruction of the lattice operators} 
We now re-write the lattice fermion operators in terms of these continuum fields. The lattice fermion operator on sub-lattice $A$ can be written as
\begin{eqnarray}
    c_{A}(\mathbf{r}) & = & \int_{\mathrm{BZ}}
\frac{d^2 k}{(2\pi)^2} \, e^{i \mathbf{k}\cdot \mathbf{r}} \, \chi_A(\mathbf{k}) \\
& \simeq &
e^{i \mathbf{k}_1\cdot \mathbf{r}} \int
\frac{d^2 q}{(2\pi)^2} \, e^{i \mathbf{q}\cdot \mathbf{r}} \, \chi_A(\mathbf{k}+\mathbf{q}),
\label{eq:lattice_to_patch}
\end{eqnarray}
where we have restricted the momentum integral to the low energy patch around $\mathbf{k}_{1}$. 
Substituting Eq.~\eqref{eq:projected_chi} into Eq.~\eqref{eq:lattice_to_patch}
and using the definition~\eqref{eq:continuum_fields}, we obtain the projected
relation between lattice and continuum operators:
\begin{eqnarray}
    c_{+\mathbf{k}_{1}, A} (\mathbf{r}) \simeq \lambda \, L^{k_1}_2(\mathbf{r})
    + \mu \, L^{k_1}_1(\mathbf{r}). \label{eq:projected_real_space}
\end{eqnarray}
The subscript $+\mathbf{k}_{1}$ in $c_{+\mathbf{k}_{1}, A} (\mathbf{r})$ reminds us that, it is a annihilation operator in a theory which is valid only around the Dirac point $+\mathbf{k}_{1}$.
Keeping this in mind, one can write down, similarly to Eq.~\eqref{c_to_l}, a relation between lattice and continuum operators as
\begin{align}
    \frac{c_{n,\alpha}}{\sqrt{\mathcal{A}}} & =  
    e^{i\mathbf{k}_1\cdot \mathbf{r}_{n}}  c_{\mathbf{k}_{1},\alpha}(\mathbf{r}_{n}) +  e^{-i\mathbf{k}_1\cdot \mathbf{r}_{n}}  c_{\mathbf{k}_{1},\alpha}^{\dagger}(\mathbf{r}_{n}) \nonumber \\
    & + e^{i\mathbf{k}_2\cdot \mathbf{r}_{n}}  c_{\mathbf{k}_{2},\alpha}(\mathbf{r}_{n}) +  e^{-i\mathbf{k}_2\cdot \mathbf{r}_{n}}  c_{\mathbf{k}_{2},\alpha}^{\dagger}(\mathbf{r}_{n}) \nonumber \\
     & \simeq e^{i \mathbf{k}_{1}\cdot \mathbf{r}_n} m_{\alpha} L_{2}^{\mathbf{k}_{1}} (\mathbf{r}_n) + e^{-i \mathbf{k}_{1}\cdot \mathbf{r}_n} m_{\alpha}^{*} L_{2}^{\mathbf{k}_{1} \dagger} (\mathbf{r}_n) \nonumber
     \\
    & + e^{i \mathbf{k}_{2}\cdot \mathbf{r}_n} n_{\alpha} L_{2}^{\mathbf{k}_{2}} (\mathbf{r}_n) + e^{-i \mathbf{k}_{2}\cdot \mathbf{r}_n} n_{\alpha}^{*} L_{2}^{\mathbf{k}_{2}\dagger} (\mathbf{r}_n) \nonumber   
\end{align}    
\begin{eqnarray}
    \frac{c_{n,\beta}}{\sqrt{\mathcal{A}}} & \simeq  e^{i \mathbf{k}_{1}\cdot \mathbf{r}_n} m_{\beta} L_{1}^{\mathbf{k}_{1}} (\mathbf{r}_n) + e^{-i \mathbf{k}_{1}\cdot \mathbf{r}_n} m_{\beta}^{*} L_{1}^{\mathbf{k}_{1} \dagger} (\mathbf{r}_n) \nonumber \\
     & + e^{i \mathbf{k}_{2}\cdot \mathbf{r}_n} n_{\beta}  L_{1}^{\mathbf{k}_{2}} (\mathbf{r}_n) + e^{-i \mathbf{k}_{2}\cdot \mathbf{r}_n} n_{\beta}^{*} L_{1}^{\mathbf{k}_{2}\dagger} (\mathbf{r}_n)  \label{small_kp_c_l}
\end{eqnarray}
where $\alpha \in \{A,B\}$, $\beta \in \{C,D\}$, $L_{1(2)}^{\mathbf{k}_{1}}(\mathbf{r}_n)$ creates a low energy (corresponding to one of the two low-energy bands) fermion  of momentum close to the Dirac point $\mathbf{k}_{1}$, at position $\mathbf{r}_n$, $\mathcal{A}=\sqrt{3}$ is the area of the unit cell, and the coefficients
\begin{eqnarray}
   m_A &=& \sqrt{\frac{3+\sqrt{3}}{6}} e^{-i \pi/4}, \quad m_C= m_A^*\nonumber \\
   m_B=m_D&=&\sqrt{\frac{3-\sqrt{3}}{6}}\nonumber \\
   n_A &=& \sqrt{\frac{3-\sqrt{3}}{6}}e^{-i 3\pi/4}, \quad n_C=n_A^*\nonumber \\
   n_B=n_D&=&\sqrt{\frac{3+\sqrt{3}}{6}}.\nonumber
\end{eqnarray}
These expressions make explicit that i) the lattice operator is a superposition of contributions from all Dirac points, ii) each contribution carries a rapid phase factor $e^{\pm i \textbf{Q}\cdot\textbf{r}}$.

One can now use equation~(\ref{small_kp_c_l}) to replace the lattice operator $c(\mathbf{r})$ with the continuum ones $L(\mathbf{r})$, in the Kitaev's Hamiltonian~(\ref{kitaevh}). Neglecting the terms proportional to $e^{i\mathbf{k}_{1(2)}\cdot\mathbf{r}}$ (fast oscillating terms), this gives us the Kitaev's Hamiltonian, in the vortex-full case, valid for low energies and in the long wavelength limit, see Eqs.~(\ref{l_basis}, \ref{kitaevhcl}, \ref{hfull}, \ref{dirack1}), written in terms of the multi-component spinor~(\ref{l_basis}).

---------------------------------------------------


%

\end{document}